\documentclass[journal]{IEEEtran}

\usepackage{amsmath,amsfonts}
\usepackage{algorithmic}
\usepackage{algorithm}
\usepackage{array}
\usepackage[caption=false,font=normalsize,labelfont=sf,textfont=sf]{subfig}
\usepackage{textcomp}
\usepackage{stfloats}
\usepackage{url}
\usepackage{verbatim}
\usepackage{graphicx}
\usepackage{cite}
\usepackage{epstopdf}
\usepackage{amssymb, bm, amsthm}
\usepackage{color, soul}
\usepackage{stfloats}
\normalsize

\ifCLASSINFOpdf
\else
\fi

\begin{document}
%
\title{{Confidential Signal Cancellation Phenomenon in Interference Alignment Networks: Cause and Cure}} 
%
%
%
\author
{
        {Lin Hu,~\IEEEmembership{{Senior Member,~IEEE,}} 
        Jiabing Fan,~\IEEEmembership{{Graduate Student Member,~IEEE,}}\\
        Hong Wen,~\IEEEmembership{{Senior Member,~IEEE,}}
        Jie Tang,~\IEEEmembership{Member,~IEEE,}
        and Qianbin Chen,~\IEEEmembership{Senior Member,~IEEE}
        }%

\thanks{This work was supported in part by the National Natural Science Foundation of China (No. 61801060),
in part by Sichuan Science and Technology Program (No. 2022YFH0098),
and in part by the National Key Research and Development Program of China (No. 2019YFB1803204 and No. 2018YFB0904905).
\emph{(Corresponding author: Hong Wen)}  } 
\thanks{{Lin Hu, Jiabing Fan, and Qianbin Chen are with the School of Communications and Information Engineering,
Chongqing University of Posts and Telecommunications, Chongqing 400065, China (e-mail: lin.hu@ieee.org; jiabing.fan.cqupt@gmail.com; chenqb@cqupt.edu.cn).}}
\thanks{{Hong Wen and Jie Tang are with the School of Aeronautics and Astronautics,
University of Electronic Science and Technology of China, Chengdu 611731, China (e-mail: sunlike@uestc.edu.cn; cs.tan@uestc.edu.cn).}}
%
%
}

\maketitle

\begin{abstract}
This paper investigates physical layer security (PLS) in wireless interference networks.
Specifically, we consider confidential transmission from a legitimate transmitter (Alice) to a legitimate receiver (Bob),
in the presence of non-colluding passive eavesdroppers (Eves), as well as multiple legitimate transceivers.
To mitigate interference at legitimate receivers and enhance PLS, artificial noise (AN) aided interference alignment (IA) is explored.
However, the conventional leakage minimization (LM) based IA may exhibit confidential signal cancellation phenomenon.
We theoretically analyze the cause and then establish a condition under which this phenomenon will occur almost surely.
Moreover, we propose a means of avoiding this phenomenon by integrating the max-eigenmode beamforming (MEB) into the traditional LM based IA.
By assuming that only statistical channel state informations (CSIs) of Eves and local CSIs of legitimate users are available,
we derive a closed form expression for the secrecy outage probability (SOP),
and establish a condition under which positive secrecy rate is achievable.
To enhance security performance,
an SOP constrained secrecy rate maximization (SRM) problem is formulated and an efficient numerical method is developed for the optimal solution.
Numerical results confirm the effectiveness and the usefulness of the proposed approach.
\end{abstract}


%
\IEEEpeerreviewmaketitle

\section{Introduction}
%
%
\IEEEPARstart{D}{ue} to the broadcasting nature,
wireless communication is vulnerable to malicious eavesdropping.
The
huge
surge in confidential data traffic (e.g., financial and military information, and personally identifiable information)
over wireless networks has made the security issue a critical concern \cite{6G_XiaohuYou}.
Traditionally, confidential issues are addressed by
complicated
cryptographic solutions with high complexity,
which have inherent difficulties in secret key distribution.
However, by exploiting only
physical properties of wireless channel, physical layer security (PLS) \cite{PHY_SUR} can secure confidential transmission against eavesdropping,
and
thus
is identified as an appealing
alternative or
complement to cryptographic techniques.

%
%
Existing works on PLS can be traced back to the wiretap channel model \cite{Wiretap_channel, Degraded, Nondegraded}.
These works demonstrate that when the quality of the main channel (from source to destination) is better than that of the eavesdropper channel (from source to eavesdropper),
the confidential message can be encoded to
achieve a low error probability at the destination,
while guaranteeing confidentiality
against the eavesdropper.
Moreover, secrecy performance is shown to increase with the quality difference between the main channel and the eavesdropper channel.
Motivated by these results, PLS has been extensively investigated in a variety of wireless scenarios,
such as the 6th generation (6G) networks \cite{6G_PHY_2},
intelligent reflecting surface (IRS) aided networks \cite{RIS},
non-orthogonal multiple access (NOMA) networks \cite{NOMA},
unmanned aerial vehicle (UAV) systems \cite{UAV},
cooperative networks \cite{Cooperative_PLS},
Internet of Things (IoT) \cite{IoT_1, IoT_2},
and satellite-terrestrial networks \cite{Satellite}, etc.

%
%
In addition to the broadcasting nature, interference is also a fundamental characteristic of wireless networks,
which is particularly detrimental to both confidentiality and capacity performance.
Therefore, interference management is a critical issue for researchers and practitioners of wireless systems.
To mitigate the impact of interference,
interference alignment (IA) constructs interference space for all interfering signals, while leaving the remaining space free from interference for the desired signal \cite{IA}.
It has been shown to be a powerful technique to achieve maximum spatial degrees of freedom (DoFs) for multiple-input multiple-output (MIMO) interference networks.

%
%
Checking the feasibility of IA for general MIMO interference channel
is NP-hard \cite{Feasibility_1, IA_NP}.
Nevertheless,
generic characteristics of wireless channel
(e.g., channel coefficients are independent and drawn from continuous distributions)
confirm the feasibility (or infeasibility) for almost all channel realizations \cite{Feasibility_2, Feasibility_3, Feasibility_4}.
Motivated by this,
several schemes (e.g., leakage minimization (LM) \cite{LM} and signal-to-interference-plus-noise-ratio maximization (Max-SINR) \cite{LM} )
can be utilized
to check the feasibility
and
obtain transceiver strategies.

\subsection{Related Works and Motivation}
%
%
Note that
the idea of IA naturally leads to the confidential data transmission
over interference networks \cite{IA_PHY_Review}.
Since no effort is spent on aligning the interference at eavesdropper,
the interference
(which can be alleviated or eliminated at legitimate receivers through IA)
would severely interfere with eavesdropper's overhearing,
and hence enhancing the confidentiality performance.
With the help of IA, the harmful multi-user interference can be employed as an efficient and effective tool for anti-eavesdropping
\cite{IA_CRN, IA_UAV, IA_AN_1, IA_AN_2, IA_AN_3, MyWork_GC}.


By assuming that perfect channel state information (CSI) of the eavesdropper is available,
zero forcing (ZF) transmission and IA are utilized to improve the secrecy rate performance for cognitive radio networks (CRNs) in \cite{IA_CRN},
and eavesdropping rate performance for UAV assisted networks \cite{IA_UAV}.


To further degrade the confidential signal reception at the eavesdropper,
artificial noise (AN) assisted IA is investigated in \cite{IA_AN_1, IA_AN_2, IA_AN_3, MyWork_GC}.
In \cite{IA_AN_1},
the power allocation ratio between the AN signal and the desired signal is predetermined instead of optimized,
leading to a suboptimal
security
performance.
The average secrecy rate maximization problem is formulated in \cite{IA_AN_2},
and a local optimal solution is obtained by convex-concave procedure \cite{convex_concave}.
The work in \cite{IA_AN_3} proposes a scheme in which receivers transmit AN signal,
and perfect CSI
is required to measure
the eavesdropping rate.
The work in \cite{MyWork_GC} considers a scenario where each transmitter sends only one data stream to its intended receiver,
which sacrifices multiplexing gain and reduces the network capacity.



Note that
the confidential signal cancellation phenomenon may occur when IA is deployed,
making PLS impractical for IA networks.
This problem is not analyzed in existing IA security approaches \cite{IA_PHY_Review, IA_CRN, IA_UAV, IA_AN_1, IA_AN_2, IA_AN_3, MyWork_GC}.
Although examples presented in \cite[Section VII]{LM}
explains the phenomenon of the desired signal cancellation, the analysis only applies to channels with specialized structure (e.g., diagonal or block diagonal forms).
Nevertheless,
to the best of our knowledge,
there is no theoretical analysis of this problem for generic channels without special structure.
In this paper,
we theoretically analyze the cause and then establish a condition under which this phenomenon will occur almost surely.
Moreover, we propose a modified LM based IA to overcome this threat,
by incorporating the max-eigenmode beamforming (MEB) for confidential transmission.

Furthermore, the works in \cite{IA_CRN, IA_UAV, IA_AN_1, IA_AN_2, IA_AN_3, MyWork_GC} assume that there exists only a single eavesdropper in IA networks,
thus limiting the application scope.
In this paper,
we consider the slow fading MIMO wiretap channel where a legitimate transmitter intends to send a confidential signal to a legitimate receiver,
in the presence of multiple passive eavesdroppers, as well as multiple legitimate transceivers.
In particular, we assume that only the statistical CSIs of eavesdroppers are available.
Under this assumption, secrecy outage may occur, which is not considered in works \cite{IA_PHY_Review, IA_CRN, IA_UAV, IA_AN_1, IA_AN_2, IA_AN_3}.
Moreover, in order to enhance PLS,
the power allocation ratio between the confidential
and  AN signals is optimized for secrecy rate maximization (SRM), subject to a maximum allowable
secrecy outage probability (SOP).

Other than IA security in \cite{IA_CRN, IA_UAV, IA_AN_1, IA_AN_2, IA_AN_3, MyWork_GC},
there are many works on PLS for cooperative networks \cite{PengchengMU_1, PengchengMU_2, MyWork_ICC, Adaptive_Cooperation, Cooperation_SRM},
Internet of Things (IoT) \cite{IoT_SOP}, cellular networks
(e.g., two-cell \cite{Two_Cell_1, Two_Cell_2, Two_Cell_3}
and multi-cell networks \cite{Multi_Cell_1, Multi_Cell_2, Multi_Cell_3}),
where interference management is also crucial to security performance.
Our work is different from these research efforts in the following aspects:
\begin{enumerate}
\item
Unlike the works in \cite{IA_CRN, IA_UAV, IA_AN_3, Two_Cell_1, Two_Cell_2}
which assume perfect CSI of the eavesdropper,
we make a more general assumption of only knowing eavesdroppers' statistical CSIs.
In addition to
the multi-user interference which is beneficial for PLS,
the AN signal is also helpful to improve security performance,
which is not considered in \cite{IA_CRN, IA_UAV, Two_Cell_1, Two_Cell_2, Multi_Cell_1}.
\item
In \cite{IA_UAV} and \cite{IA_AN_2}, the eavesdropping rate is utilized as the secrecy assessment,
which is more suited for legitimate proactive instead of illegitimate passive eavesdropping scenario \cite{Proactive_Eavesdropping}.
Besides, the ergodic secrecy rate (ESR) is used for the performance metric in \cite{IA_AN_1, IA_AN_2, Multi_Cell_1, Multi_Cell_2, Multi_Cell_3}.
This metric is more appropriate to fast fading channels, in which the codewords spread over a wide range of channel variations.
In this paper, we
focus
on the SOP constrained secrecy rate, which is more suitable for slow fading channels or delay-limited applications.
\item
For cooperative networks,
the SOP constrained SRM problem is investigated
in \cite{PengchengMU_1, PengchengMU_2, MyWork_ICC}.
Based on \cite{MyWork_ICC},
adaptive jamming for secure communication is designed for cooperative \cite{Adaptive_Cooperation} and two-cell networks \cite{Two_Cell_3}.
In addition, cooperative jamming is studied for SRM in \cite{Cooperation_SRM} and for SOP minimization in \cite{IoT_SOP}.
Note that all these works do not consider multi-user interference mitigation, and hence cannot be directly applied to IA networks.
In \cite{Cooperation_SRM},
the SOP is evaluated by taking advantage of properties of confluent and Gauss hypergeometric functions.
However,
it is difficult to use this approach to obtain a closed form of SOP for IA networks.
To circumvent this difficulty, we resort to the Laplace transform \cite[12.11]{Table},
which is able to recover the complementary cumulative distribution function (CCDF) of a continuous random variable.
Surprisingly, the SOP for IA networks can be calculated precisely using Laplace transform analysis.

\end{enumerate}

\subsection{Our Contributions and Paper Organization}
The main contributions can be summarized as follows:
\begin{enumerate}
\item
We theoretically analyze the cause of confidential signal cancellation phenomenon in interference networks,
which is not considered in previous IA security schemes.
Furthermore, we establish a condition under which this phenomenon will occur almost surely.
Numerical results are consistent with the theoretical analysis.
\item
To overcome confidential signal cancellation, we propose a modified LM based IA, by incorporating the MEB for confidential transmission.
The feasibility of IA is also analyzed, and guiding insight is provided into the practical system designs,
such as the guidelines for the selection of the dimension of AN.
Numerical results demonstrate that our scheme is practical and reliable.
\item
To enhance security performance,
we formulate a SOP constrained SRM problem.
We first present an accurate closed form expression for SOP using Laplace transform analysis,
and establish the condition under which positive secrecy rate is achievable.
The explicit transmit design is obtained by optimizing power allocation ratio between confidential signal and AN signal.
Numerical results confirm the effectiveness of our approach.


%
\end{enumerate}

The rest of this paper is organized as follows.
In Section II, system model and IA problem are described.
In Section III,
after analyzing the cause of confidential signal cancellation,
a modified IA is proposed to remedy this problem.
The optimal transmit design for SOP constrained SRM is provided in Section IV,
and we conclude this paper in Section V.


\emph{Notations:}
$\mathbb{C}^n$ represents the $n$-dimensional complex space, and ${\mathbb{C}^{M\!\times\!N}}$ represents the space of complex $M\!\times\!N$ matrices.
$\left(\cdot\right)^H$ and $\prod$ denote Hermitian transpose and product operators, respectively.
$\left\{\mathbf{X}_i\right\}_{i=1}^{K}$ is the horizontal concatenation of $\mathbf{X}_1,\ldots,\mathbf{X}_K$.
$\text{span}\left(\mathbf{X}\right)$ and $\text{null}\left(\mathbf{X}\right)$ denote
the column and null spaces of $\mathbf{X}$, respectively.
$\dim[\cdot]$ denotes the dimension of a subspace.
%
%
$\mathcal{CN}\left(\bm{\mu},\mathbf{Q}\right)$ is
the circularly symmetric complex Gaussian distribution with mean $\bm{\mu}$ and covariance $\mathbf{Q}$.
$\Gamma \left(\alpha,\lambda\right)$ is the gamma distribution with shape $\alpha$ and rate $\lambda$,
and $\text{Exp}\left(\lambda\right)$ is the exponential distribution with rate $\lambda$.
$\Pr\left\{\cdot\right\}$ denotes the probability operation.
$\lfloor x\rfloor$, $\lceil x\rceil$, and $\vert\cdot\vert$ denote floor, ceil, and modulus functions, respectively.
$\Vert\cdot\Vert$ and $\Vert\cdot\Vert_F$ are $\ell_2$ and Frobenius norms, respectively.
$\log \left(\cdot\right)$ and $\ln \left(\cdot\right)$ denote base-2 and natural logarithms, respectively.
The Laplace transform of a function $f\left(t\right)$, defined for real numbers $t\!\ge\!0$, is denoted by
$L\left[f\left(t\right);s\right]\!=\!\int_{0}^{\infty}{f\left(t\right)e^{-st}}dt$ \cite[12.11]{Table}.


\section{System Model and IA Problem}
We consider the system model as shown in Fig. 1,
where a legitimate source (Alice) transmits a confidential message to a legitimate destination (Bob),
in the presence of $L$ non-colluding passive eavesdroppers (Eves) trying to decode this message.
In addition,
there exist $K$ legitimate transmitters that transmit general messages to their intended receivers.
The set of transceivers is defined as $\mathcal{K} \! \triangleq \! \{1,2,\ldots,K\}$,
and the set of Eves is defined as $\mathcal{L} \! \triangleq \! \{1,2,\ldots,L\}$.
Note that Alice and Tx $k$ are legitimate transmitters, while Bob and Rx $k$ are legitimate receivers, for all $k\!\in\!\mathcal{K}$.
We assume that Alice and Bob are equipped with $M_a$ and $N_b$ antennas, respectively,
each Eve is equipped with a single antenna,
while Tx $k$ and Rx $k$ are equipped with $M_k$ and $N_k$ antennas, respectively.

All channels are assumed to be generic under slow flat fading,
and the CSIs are invariant within one coherence time block but independently changes from one block to another.
Channels from Alice to Bob, Rx $j$, and Eve $l$ are
$\mathbf{H}_{ba}\!\in\!\mathbb{C}^{{N_b}\!\times\!{M_a}}$,
$\mathbf{H}_{ja}\!\in\!\mathbb{C}^{{N_j}\!\times\!{M_a}}$,
and $\mathbf{h}_{{e_l},a}\!\in\!\mathbb{C}^{M_a}$, respectively,
and channels from Tx $k$ to Bob, Rx $j$, and Eve $l$ are
$\mathbf{H}_{bk}\!\in\!\mathbb{C}^{{N_b}\!\times\!{M_k}}$,
$\mathbf{H}_{jk}\!\in\!\mathbb{C}^{{N_j}\!\times\!{M_k}}$,
and $\mathbf{h}_{{e_l},k}\!\in\!\mathbb{C}^{M_k}$, respectively,
$\forall l\!\in\!\mathcal{L}$, $\forall j,k\!\in\!\mathcal{K}$.
Eves' statistical CSIs are assume to be available,
and $\mathbf{h}_{{e_l},a}\!\sim\!\mathcal{CN}\left(\mathbf{0},\mathbf{I}_{M_a}\right)$,
$\mathbf{h}_{{e_l},k}\!\sim\!\mathcal{CN}\left(\mathbf{0},\mathbf{I}_{M_k}\right)$.
For the purpose of distributed implementation of IA,
we assume that local CSIs of legitimate users are available,
which means that each legitimate receiver can accurately estimate channels between itself and legitimate transmitters through pilot signaling,
and each legitimate transmitter can obtain the CSI of its associated legitimate receivers through uplink feedback.
Similar assumptions can be found in \cite{Two_Cell_3, Local_CSI_2, Local_CSI_3}.

For confidential transmission from Alice to Bob, the AN aided beamforming is performed.
Let $\mathbf{v}_a\!\in\!\mathbb{C}^{{M_a}}$ and $\mathbf{W}_a\!\in\!\mathbb{C}^{{M_a}\!\times\!{d_{a}}}$
be secrecy beamforming vector and unitary precoding matrix,
respectively,
where $d_{a}\!\in\!\left[1, M_a\!-\!1\right]$ denotes the dimension of AN.
Let $P_a$ be the transmit power of Alice,
and let $\theta\!\in\!\left[ 0,1 \right]$ be the fraction of $P_a$ allocated to the confidential signal.
Then the transmitted signal from Alice is given by
%
%
\begin{align}
\mathbf{s}_a
\!=\!\sqrt{\theta P_a }{\mathbf{v}_a}{x_a}
\!+\!\sqrt{{\left(1\!-\!\theta\right){P_a}}/{d_a}} {\mathbf{W}_a}{\mathbf{z}_a},\tag{1}
\end{align}
where
$x_a\!\sim\!\mathcal{CN}\left(0,1\right)$ is the confidential data symbol for Bob,
and ${\mathbf{z}_a}\!\sim\!\mathcal{CN}\left( \mathbf{0},\mathbf{I}_{d_{a}}\right)$ denotes the Gaussian noise vector.

\begin{figure}[t]
\centering
\includegraphics[height=6.26cm]{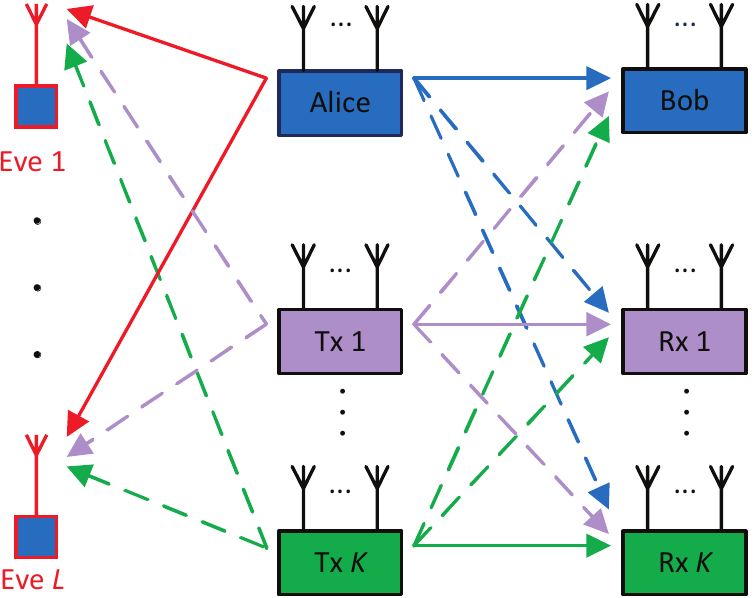}
\caption{The multi-user interference networks in the presence of non-colluding passive eavesdroppers.}
\label{Fig. 1}
\end{figure}

Note that the first term on the right-hand side of (1) is confidential signal,
and the second term is AN signal.
Besides, $\theta\!=\!1$ corresponds to the secrecy beamforming without AN,
and $\theta\!=\!0$ means that the secure transmission is suspended.


For the general message transmission,
Tx $k$ sends $d_k$ data
streams to Rx $k$ with transmit power $P_k$, $\forall k\!\in\!\mathcal{K}$.
%
%
In order to achieve IA, Rx $k$ preserves at least a one-dimensional subspace for interference and AN,
and thus $1\!\le\!d_k\!\le\!\min\left(M_k,N_k\!-\!1\right)$.
Let ${\mathbf{x}_k}\!\sim\!\mathcal{CN}\left(\mathbf{0},\mathbf{I}_{d_k}\right)$
and $\mathbf{V}_k\!\in\!\mathbb{C}^{{M_k}\!\times\!{d_k}}$
be data symbol vector for Rx $k$ and unitary precoding matrix at Tx $k$, respectively.
Then the signal transmitted by Tx $k$ can be expressed as
%
%
\begin{align}
\mathbf{s}_k \!=\! \sqrt{P_k/{d_k}}{\mathbf{V}_k}{\mathbf{x}_k}.\tag{2}
\end{align}

Let $\mathbf{u}_b\!\in\!\mathbb{C}^{{N_b}}$ denote the receive combining vector at Bob,
and let $\mathbf{U}_k\!\in\!\mathbb{C}^{{N_k}\!\times\!{d_k}}$ denote the unitary combining matrix at Rx $k$,
$\forall k\!\in\!\mathcal{K}$.
After the receive combining, the obtained signals at Bob, Rx $k$, and Eve $l$ can be expressed as follows:
%
%
\begin{align}
y_b\!
  =&
\sqrt{\theta P_a }\mathbf{u}_{b}^{H} {\mathbf{H}_{ba}} {\mathbf{v}_a}{x_a} \!+\!\sqrt{ {\left( 1\!-\!\theta \right){P_a}}/{d_{a}} } \mathbf{u}_{b}^{H} {\mathbf{H}_{ba}} {\mathbf{W}_a} {\mathbf{z}_a} \nonumber\\
   &
+\!\mathbf{u}_{b}^{H}\sum\nolimits_{k=1}^{K}{\sqrt{P_k/d_k} {\mathbf{H}_{bk}} {\mathbf{V}_k} {\mathbf{x}_k}} \!+\!\mathbf{u}_{b}^{H} {\mathbf{n}_b} , \tag{3} \\
\mathbf{y}_k\!
  =&
   { {\sqrt{ \theta P_a}} \mathbf{U}_{k}^{H} \mathbf{H}_{ka} {\mathbf{v}_a} {x_a}  }
   \!+\!{ \sqrt{ {\left( 1\!-\!\theta \right){P_a}}/{d_{a}} } {\mathbf{U}_{k}^H}\mathbf{H}_{ka}{\mathbf{W}_a}{\mathbf{z}_a} }\nonumber \\
   &
   +\!{\mathbf{U}_{k}^{H} \sum\nolimits_{j=1}^{K}\sqrt{P_j/d_j}\mathbf{H}_{kj}{\mathbf{V}_j}{\mathbf{x}_j} \!+\! \mathbf{U}_{k}^{H}\mathbf{n}_k}, \forall k\!\in\!\mathcal{K},\tag{4}\\
y_{e_l}\!
  =&
   {\sqrt{ \theta P_a}}\mathbf{h}_{{e_l},a}^H {\mathbf{v}_a} {x_a}
   \!+\! {\sqrt{ {\left( 1\!-\!\theta \right){P_a}}/{d_{a}} } \mathbf{h}_{{e_l},a}^H{\mathbf{W}_a}{\mathbf{z}_a}  }\nonumber\\
   &
   +\! \sum\nolimits_{k=1}^{K}\sqrt{P_k/d_k}{\mathbf{h}_{{e_l},k}^{H} {\mathbf{V}_k}{\mathbf{x}_k} }+{{n}_{e_l}}, \forall l\!\in\!\mathcal{L}, \tag{5}
\end{align} where
$\mathbf{n}_b\!\sim\!\mathcal{CN}\left( \mathbf{0},{{\mathbf{I}}_{N_b}}\right)$,
${\mathbf{n}_k}\!\sim\!\mathcal{CN}\left(\mathbf{0}, {\mathbf{I}_{N_k}}\right)$,
${n_{e_l}}\!\sim\!\mathcal{CN}\left( 0,1 \right)$
are independent
of each other,
and represent additive complex white Gaussian noises (AWGNs) at Bob, Rx $k$, and Eve $l$, respectively.
We adhere to the notation of \cite{Feasibility_1},
and denote the system model as
$\mathcal{X}\!=\!\left(M_a\!\times\!N_b,\left[1,d_{a}\right]\right)\prod_{k=1}^{K}\left(M_k\!\times\!N_k,d_k\right)$.

\emph{Remark 2.1:} 
Note that the AN signal transmitted by Alice and signals transmitted by $\text{Tx } 1,\ldots, \text{Tx }K$
will interfere with Eves.
In addition, Alice and other transmitters will interfere with each other,
reducing confidential signal quality at Bob and desired signal quality at $\text{Rx }1,\ldots,\text{Rx }K$.
Thus, transceiver strategies need to be carefully designed to mitigate multi-user interference and enhance security performance.

We
concentrate
on alignment in spatial dimension by using multiple antennas.
The original IA problem is equivalent to design transceivers such that
1) both AN and all interfering signals at each legitimate receiver fall into a reduced dimensional subspace
(also called the \emph{interference subspace}),
and 2) the desired signal at each legitimate receiver is not within the interference subspace.
Hence, the desired signal can be extracted by zero forcing (ZF) both AN and interfering signals.
To this end,  the following conditions must be satisfied:
%
%
\begin{align}
&\left[\mathbf{H}_{ba}^H\mathbf{u}_b, \left\{\mathbf{H}_{ka}^H\mathbf{U}_{k}\right\}_{k=1}^K\right]^H\mathbf{W}_a \!=\! \mathbf{0}, \tag{6}\\
&\left[  \left\{\mathbf{H}_{ka}^H\mathbf{U}_{k}\right\}_{k=1}^K  \right]^H\mathbf{v}_a\!=\!\mathbf{0}, \tag{7}\\
&\left[\mathbf{H}_{bk}^H\mathbf{u}_b, \left\{\mathbf{H}_{jk}^H\mathbf{U}_j\right\}_{j=1,j\ne k}^K\right]^H\mathbf{V}_k\!=\!\mathbf{0},\,  k\!=\!1,\ldots,K, \tag{8}\\
&\;\mathbf{u}_{b}^{H}\mathbf{H}_{ba}\mathbf{v}_a\!\ne\!0, \text{or equivalently}, \text{rank}\left(\mathbf{u}_{b}^{H}\mathbf{H}_{ba}\mathbf{v}_a\right)\!=\!1, \tag{9} \\
&\;\text{rank} \left(\mathbf{U}_{k}^{H}\mathbf{H}_{kk}\mathbf{V}_{k}\right) \!=\! d_k,\, k\!=\!1,\ldots,K.\tag{10}
\end{align}
where (6)--(8) require that both AN and multi-user interference are eliminated at each legitimate receiver,
and (9)--(10) require that the confidential signal and other desired signals are visible and resolvable within the AN and interference free subspace.

\emph{Remark 2.2:}
Note that the price paid for IA is that the desired signal may not be exactly orthogonal to the interference subspace.
The gain is that the confidential signal and other desired signals will not be interfered by AN and multi-user interference,
leading to an optimal design for network capacity maximization at high signal-to-noise ratio (SNR) regime.


\emph{Remark 2.3:}
Note that (6)--(8) can be interpreted as a system of polynomial equations with
$\mathbf{v}_a$, $\mathbf{W}_a$, $\mathbf{u}_b$, $\mathbf{V}_k$, and $\mathbf{U}_k$, $k\!=\!1,\ldots,K$,
being the variables.
Specifically, the traditional iterative alignment approach (e.g., the LM based IA) is
driven by the aim of finding nontrivial solution to (6)--(8),
with little regard to the fate of the desired signal at the desired receiver,
which corresponds to conditions (9)--(10).
Suppose that the solution to (6)--(8) is obtained.
If the obtained solution does not satisfy (9), then the confidential signal cancellation occurs at Bob,
and if it does not satisfy (10), then the desired signal cancellation occurs at other legitimate receivers.

\emph{Remark 2.4:} 
Although unitariness is not a necessary requirement for IA equations (6)--(8),
we assume for simplicity that $\mathbf{V}_k$ and $\mathbf{U}_k$ are unitary, $\forall k\!\in\!\mathcal{K}$.
Note that unitary transmit precoding and receive combining does not affect the feasibility of (6)--(8),
and this property can be used to simplify the design of iterative algorithms suitable for distributed implementation \cite{LM}.
Unitariness is also desired in MIMO precoding designs to aid with feedback of channel state.
In addition, it allows us to extend interference channel to interference broadcasting channel,
e.g., Tx $k$ serves $d_k$ users, and the $i$-th column of $\mathbf{V}_k$ denotes a beamforming vector for user $i$, $i\!=\!1,\ldots, d_k$.

\section{Analysis of \\the Traditional LM based IA and the Modified IA}

In this section, we first carry out an analysis on the traditional LM based IA in terms of its complexity, feasibility,
and the reason of confidential signal cancellation.
Then a modified IA scheme is proposed for combatting signal cancellation, by integrating the MEB
into the traditional LM based method.

The analysis framework presented in this paper is based on the following assumption:
\begin{align}
  M_a \!\ge\! 1\!+\! \sum\nolimits_{k=1}^{K}\!d_k,\, M_k \!\ge\! 1\!+\! \sum\nolimits_{k=1}^{K}\!d_k,\, k\!=\!1,\ldots,K.\tag{11}
\end{align}
Note that
our analysis framework is based on assumption (11).
In particular,
under the assumption that $M_a \!\ge\! 1\!+\! \sum\nolimits_{k=1}^{K}\!d_k$,
we theoretically prove that the confidential signal cancellation phenomenon will occur under a certain condition on the dimension of AN (see Section III-A.3 for details).

\emph{Remark 3.1:}
Assumption (11) implies that each legitimate transmitter is equipped with a sufficient number of antennas.
Note that upscaling the number of antennas at transmitters is a key characteristic of current and future wireless communication systems \cite{Massive_MIMO}.
Thus (11) is reasonable and practical for multi-user interference networks,
and could further harvest the multiplexing and array gains of large number of antennas.

\emph{Remark 3.2:}
It may be appropriate to assume (11) when the IA in signal space is considered for interference networks.
Note that more antennas can create more spatial dimensions, which is instrumental to the development of IA,
i.e., consolidating AN and interfering signals into the interference subspace.
As will be discussed in Section III-A.1, (11) is also helpful in
simplifying the IA problem.
Then the mechanism behind the confidential signal cancellation phenomenon can be explained in a rigorous and tractable manner.
However, a comprehensive mathematical analysis of this phenomenon for the general scenario without assuming (11) turns out to be a nontrivial task,
and will be considered in future work.

To simplify notations, we define the following matrices:
\begin{align}
\mathbf{M}& \!\triangleq\! \left[\mathbf{H}_{ba}^H\mathbf{u}_b, \left\{\mathbf{H}_{ka}^H\mathbf{U}_{k}\right\}_{k=1}^K\right]^H,\tag{12}\\
\bar{\mathbf{M}}&\!\triangleq\! \left[  \left\{\mathbf{H}_{ka}^H\mathbf{U}_{k}\right\}_{k=1}^K  \right]^H,\tag{13}\\
\mathbf{M}_k&   \!\triangleq\! \left[\mathbf{H}_{bk}^H\mathbf{u}_b, \left\{\mathbf{H}_{jk}^H\mathbf{U}_j\right\}_{j=1,j\ne k}^K\right]^H,\,  k\!=\!1,\ldots,K. \tag{14}
\end{align}
Note that $\mathbf{M}$ and $\bar{\mathbf{M}}$ are matrices of size
$\left(1\!+\! \sum\nolimits_{k=1}^{K}\!d_k\right)\!\times\!M_a$ and $\left(\sum\nolimits_{k=1}^{K}\!d_k\right)\!\times\!M_a$, respectively,
and $\mathbf{M}_k$, $k\!=\!1,\ldots,K$, are matrices of size $\left(1\!+\! \sum\nolimits_{j=1,j\neq k}^{K}\!d_j\right)\!\times\!M_k$.
Then we have that
\begin{align}
\text{rank}\left(\mathbf{M}\right)        &\!\le\!    1\!+\! \sum\nolimits_{k=1}^{K}\!d_k         \!\le\!     M_a,\tag{15}\\
\text{rank}\left(\bar{\mathbf{M}}\right)  &\!\le\!    \sum\nolimits_{k=1}^{K}\!d_k                \!\le\!     M_a\!-\!1,\tag{16}\\
\text{rank}\left(\mathbf{M}_k\right)      &\!\le\!    1\!+\! \sum\nolimits_{j=1,j\neq k}^{K}\!d_j \!\le\!     M_k\!-\!d_k.   \tag{17}
\end{align}
Thus by (16)--(17) we have the following lemma.

\emph{Lemma 3.1:} Under (11), we have the following result.
\begin{align}
\!\!\!  \dim\left[\text{null}\left(\bar{\mathbf{M}}\right)\right]    \!\ge\!1,
  \dim\left[\text{null}\left(\mathbf{M}_k\right)\right]        \!\ge\!d_k,\,  k\!=\!1,\ldots,K.    \tag{18}
\end{align}

\begin{IEEEproof}
According to \cite{Matrix}, we obtain that
\begin{align}
  &\dim\left[\text{null}\left(\bar{\mathbf{M}}\right)\right]
  \!=\!M_a\!-\!\text{rank}\left(\bar{\mathbf{M}}\right),  \tag{19}\\
  &\dim\left[\text{null}\left(\mathbf{M}_k\right)\right]\!=\!M_k\!-\!\text{rank}\left(\mathbf{M}_k\right),\,  k\!=\!1,\ldots,K. \tag{20}
\end{align}
Then the desired result directly follows from (16)--(17).
\end{IEEEproof}

Note that (18) also implies that there exist signal spaces with sufficient dimensions
that can be exploited by legitimate transmitters for interference free transmission.




\subsection{Analysis of the Traditional LM based IA}
Different from the original IA problem (6)--(10),
the traditional LM based IA problem amounts to finding a nontrivial solution to
the system of polynomial equations
(6)–(8),
without considering the rank related conditions (9)--(10).
To solve (6)--(8), we can formulate the following optimization problem:
\begin{align}
  &\mathop{\min}_{\mathbf{W}_a, \mathbf{v}_a, \mathbf{u}_b, \mathbf{V}_k, \mathbf{U}_k}
  \Vert{\mathbf{M}\mathbf{W}_a}\Vert_F^{2}\!+\!\Vert{\bar{\mathbf{M}}\mathbf{v}_a}\Vert^{2}\!+\!\sum\nolimits_{k=1}^{K}\Vert{\mathbf{M}_k\mathbf{V}_k}\Vert_F^{2}\nonumber\\
  &\qquad\ \, \text{s.t.}\qquad\quad \mathbf{W}_a^H\mathbf{W}_a\!=\!\mathbf{I}_{d_a};\, \Vert\mathbf{v}_a\Vert\!=\!\Vert\mathbf{u}_b\Vert\!=\!1;\nonumber\\
   &\qquad\qquad\qquad\ \ \mathbf{V}_k^H\mathbf{V}_k\!=\!\mathbf{U}_k^H\mathbf{U}_k\!=\!\mathbf{I}_{d_k},\,  k\!=\!1,\ldots,K,\tag{21}
\end{align}
where $\mathbf{M}$, $\bar{\mathbf{M}}$, and $\mathbf{M}_k$ are defined in (12)--(14), respectively.

Note that the objective function of (21) can be interpreted as the total interference leakage at all legitimate receivers.
In addition, the optimization variables can be partitioned into two categories:
transmit filters ($\mathbf{W}_a$, $\mathbf{v}_a$, and $\mathbf{V}_k$, $k\!=\!1,\ldots,K$) and receive filters ($\mathbf{u}_b$ and $\mathbf{U}_k$, $k\!=\!1,\ldots,K$).
Problem (21) can be solved by LM based IA which alternatively optimizes over the transmit and receive filters.
For the detailed process of this method,
interested readers may refer to \cite{LM}. 

However,
under
(11) and with
Lemma 3.1,
the traditional LM based solution process can be simplified considerably.

First, note that the term
$\Vert{\bar{\mathbf{M}}\mathbf{v}_a}\Vert^{2}\!+\!\sum\nolimits_{k=1}^{K}\Vert{\mathbf{M}_k\mathbf{V}_k}\Vert_F^{2}$ in the objective function of (21)
is lower bounded by zero.
Moreover, for any receive filters $\mathbf{u}_b$ and $\mathbf{U}_k$, $k\!=\!1,\ldots,K$,
it follows by Lemma 3.1 that there always exist $\mathbf{v}_a$ and $\mathbf{V}_k$,
such that this lower bound is attained, meaning that equations ${\bar{\mathbf{M}}\mathbf{v}_a}\!=\!\mathbf{0}$ and ${\mathbf{M}_k\mathbf{V}_k}\!=\!\mathbf{0}$, $k\!=\!1,\ldots,K$, hold trivially.
%
%
Hence, the LM based IA equations (6)--(8) can be reduced to (6).
To solve (6), we can formulate the following optimization problem:
\begin{align}
  &\!\mathop{\min}_{\mathbf{W}_a,\mathbf{u}_b,\mathbf{U}_k}\!\!\!\!
  \Vert{\mathbf{M}\mathbf{W}_a}\Vert_F^{2}\!=\!\!\Vert{\mathbf{u}_b^H\mathbf{H}_{ba}\mathbf{W}_a\Vert}^2\!\!\!+\!\!\sum\nolimits_{k=1}^{K}\!\!\Vert{\mathbf{U}_k^H\mathbf{H}_{ka}\mathbf{W}_a\Vert}_F^2\nonumber\\
  &\!\!\!\text{s.t.}\,\mathbf{W}_a^H\mathbf{W}_a\!=\!\mathbf{I}_{d_a}; \Vert\mathbf{u}_b\Vert\!=\!1; \mathbf{U}_k^H\mathbf{U}_k\!=\!\mathbf{I}_{d_k}, k\!=\!1,\ldots,K.\tag{22}
\end{align}

\emph{Remark 3.3:}
Note that problem (22) can also be considered as a MIMO downlink interference mitigation problem.
The corresponding system model is shown in Fig. 2, where a base station (Alice) only transmits the AN signal.
The objective is to design $\mathbf{W}_a$, $\mathbf{u}_b$, and $\mathbf{U}_k$, $k\!=\!1,\ldots,K$, to mitigate the interference caused by AN at all legitimate receivers.

Note that problems (22) and (21) are equivalent, and the alternating optimization technique \cite{NLP} is also applicable for solving (22), where two cases are characterized as follows.

\begin{enumerate}
  \item Case 1: $M_a\!\ge\!1\!+\!d_a\!+\!\sum\nolimits_{k=1}^{K}d_k$.\\
By \cite{Matrix} and (15), we have that $\dim\left[\text{null}\left(\mathbf{M}\right)\right]\!\ge\!d_a$,
and thus for any $\mathbf{u}_b$ and $\mathbf{U}_k$, $k\!=\!1,\ldots,K$,
simply ZF suffices to achieve the optimum of (22).

  \item Case 2: $1\!+\!\sum\nolimits_{k=1}^{K}d_k\!\le\!M_a\!\le\!d_a\!+\!\sum\nolimits_{k=1}^{K}d_k$.\\
In this case, the LM based scheme can be employed to alternatively optimize over transmit filter ($\mathbf{W}_a$) and receive filters ($\mathbf{u}_b$ and $\mathbf{U}_k$, $k\!=\!1,\ldots,K$).
\end{enumerate}
The execution process is summarized in Algorithm 1.

Note that the
iterative
solution procedure for problem (22) circumvents the need for updating
transmit filters
$\mathbf{v}_a$ and $\mathbf{V}_k$, $k\!=\!1,\ldots,K$, as compared to problem (21).
In particular,
in the last step of Algorithm 1,
the corresponding 
transmit filters
$\mathbf{v}_a$ and $\mathbf{V}_k$, can be obtained trivially via ZF method.


\begin{figure}[ht]
\centering
\includegraphics[height=5.27cm]{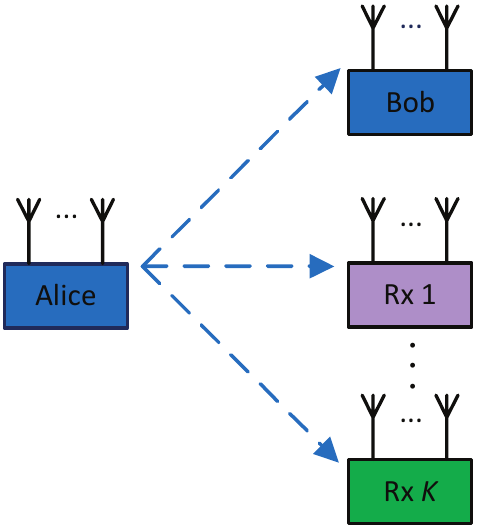}
\caption{The equivalent MIMO downlink interference mitigation scenario.}
\label{Fig. 2}
\end{figure}


%
%

%
%
\begin{algorithm}[ht]\label{Algorithm 1}
\caption{Solving problem (22) under assumption (11)}
\begin{enumerate}
\item
Initialization: Given local CSI at each legitimate
transmitter and receiver.
\item
Randomly generate $\mathbf{u}_b$ and $\mathbf{U}_k$, $k\!=\!1,\ldots,K$.
%
\item \textbf{if} $M_a\!\ge\!1\!+\!d_a\!+\!\sum\nolimits_{k=1}^{K}d_k$, \textbf{then}\\
Set the columns of $\mathbf{W}_a$ to be an orthonormal basis for $\text{null}\left(\mathbf{M}\right)$, and go to 7).

\item\textbf{else repeat}

\begin{enumerate}

\item

Choose the columns of $\mathbf{W}_a$ as the $d_a$  least dominant eigenvectors of $\mathbf{M}^H\mathbf{M}$.
    \item
Set $\mathbf{u}_b$ to be the least dominant eigenvector of $\mathbf{H}_{ba}\mathbf{W}_a\mathbf{W}_a^H\mathbf{H}_{ba}^H$,
and set columns of $\mathbf{U}_k$ as the $d_k$ least dominant eigenvectors of $\mathbf{H}_{ka}\mathbf{W}_a\mathbf{W}_a^H\mathbf{H}_{ka}^H$.
\end{enumerate}
\item \textbf{until} the objective value of (22) converges or maximum number of iterations is reached.

\item \textbf{end if}
\item
Let $\mathbf{0}\!\neq\!\mathbf{v}_a\!\in\!\text{null}\left(\bar{\mathbf{M}}\right)$, and let columns of $\mathbf{V}_k$ be an orthonormal basis for $\text{null}\left(\mathbf{M}_k\right)$, $k\!=\!1,\ldots,K$.
\end{enumerate}
\end{algorithm}

Since the objective value of (22) decreases with iteration 
and the objective function is lower bounded by zero, it follows that Algorithm 1 converges.
However, the global optimum cannot be guaranteed due to the nonconvexity of (22).

\emph{Remark 3.4:}
Note that $\mathbf{v}_a$ obtained by the last step of Algorithm 1 can be
refined to improve the confidential signal quality,
which is denoted by the main channel gain $\vert\mathbf{u}_b^H\mathbf{H}_{ba}\mathbf{v}_a\vert^2$.
The corresponding optimization problem is formulated as:
\begin{align}
\mathop{\max}_{\mathbf{v}_a} \vert\mathbf{u}_b^H\mathbf{H}_{ba}\mathbf{v}_a\vert^2\quad\text{s.t.}\ \mathbf{v}_a\!\in\!\text{null}\left(\bar{\mathbf{M}}\right).\tag{23}
\end{align}
Assume that $\mathbf{N}\!\in\!\mathbb{C}^{M_a\!\times\!\left(M_a\!-\text{rank}\left(\bar{\mathbf{M}}\right)\right)} $ is an orthonormal basis of $\text{null}\left(\bar{\mathbf{M}}\right)$
and $\mathbf{u}_b^H\mathbf{H}_{ba}\mathbf{N}\!\ne\!\mathbf{0}$.
Thus the optimal solution
is given by
$\mathbf{v}_a\!=\!\mathbf{N}\mathbf{N}^H\mathbf{H}_{ba}^H\mathbf{u}_b/\Vert\mathbf{N}\mathbf{N}^H\mathbf{H}_{ba}^H\mathbf{u}_b\Vert$.
If $\mathbf{u}_b^H\mathbf{H}_{ba}\mathbf{N}\!=\!\mathbf{0}$, then we have $\mathbf{u}_b^H\mathbf{H}_{ba}\mathbf{v}_a\!=\!0$, resulting in the confidential signal cancellation.
This problem will be analyzed in Section III-A.3.

\emph{Remark 3.5:}
Under (11), the LM based IA equations are reduced from (6)--(8) to (6). To solve (6), we formulate an optimization model (22) and develop Algorithm 1.
However, the equivalence between (22) and (6) may not hold.
If the objective value of (22) is zero, then the solution obtained with Algorithm 1 also satisfies equation (6).
Otherwise, (6) is infeasible almost surely due to the generic channel assumption.

%
%
\begin{table}[t]
  \centering
  \caption{The Complexity of Several Operations.}
  \begin{tabular}{|c|c|}
  \hline
  $\Vert\mathbf{a}\Vert$                & $4m$                 \\ \hline
  $\mathbf{A}\!+\!\mathbf{B}$           & $2mn$                \\ \hline
  $\Vert\mathbf{A}\Vert_F$              & $4mn$                \\ \hline
  $\mathbf{A}\mathbf{C}$                & $\left(8n\!-\!2\right)mp$       \\ \hline
  $\mathbf{A}\mathbf{A}^H$              & $\left(4n\!-\!1\right)m^2$       \\ \hline
  $\text{SVD}\left(\mathbf{A}\right)$   & $24mn^2\!+\!48m^2n\!+\!54m^3$            \\ \hline
\end{tabular}
\end{table}

\subsection*{A.1  Complexity Analysis of Algorithm 1}
In this paper, the computational complexity is measured by the number of real floating point operations (flops) \cite{Matrix}.
Each real addition, subtraction, multiplication, division, or square root counts as one flop.
For example, one complex addition and multiplication
are counted as
two and six flops, respectively.
Let $\mathbf{a}\!\in\!\mathbb{C}^m$, $\mathbf{A}$, $\mathbf{B}\!\in\!\mathbb{C}^{m\!\times\!n}$, $m\!\le\!n$,
and let $\mathbf{C}\!\in\!\mathbb{C}^{n\!\times\!p}$.
The flop counts for several matrix operations are summarized in Table I,
where $\text{SVD}\left(\cdot\right)$ denotes the singular value decomposition.

Note that for the nontrivial case, only $\mathbf{W}_a$, $\mathbf{u}_b$, and $\mathbf{U}_k$, $k\!=\!1,\ldots,K$, are involved in the iteration of Algorithm 1.
The required number of flops can be calculated as follows:
\begin{align}           
\!\text{flops}\left(\mathbf{W}_a\right)\!=&126M_a^3\!+\!M_a^2\left(4\sum\nolimits_{k=1}^K d_k\!+\!3\right)\nonumber\\
                                        &+\!M_a\left(\sum\nolimits_{k=1}^{K}\left(8N_k-2\right)d_k \!+\!8N_b\!-\!2\right),\tag{24}\\
                                        \!\text{flops}\left(\mathbf{u}_b\right)\!=&126N_b^3\!+\!N_b^2\left(4d_a\!-\!1\right)\!+\!N_bd_a\left(8M_a\!-\!2\right),\tag{25}\\
\!\text{flops}\left(\mathbf{U}_k\right)\!=&126N_k^3\!+\!N_k^2\left(4d_a\!-\!1\right)\!+\!N_kd_a\left(8M_a\!-\!2\right).\tag{26}
\end{align}

To simplify the analysis, we define the following notations:
\begin{align}
  S\!\triangleq\!\max\nolimits_{1\le k \le K}\max\left(M_k,N_k\right);\, T\!\triangleq\!\max(M_a,N_b).\tag{27}
\end{align}
Then we have that
\begin{align}
d_k\!\le\!\min\left(M_k,N_k\right)\!\le\!S, \forall k\!\in\!\mathcal{K};\, 1\!\le\!d_{a}\!\le\!M_a\!-\!1\!<\!T. \tag{28}
\end{align}
For the nontrivial case, the number of flops required for each iteration of Algorithm 1 can be characterized as
\begin{align}
\text{Flops}
\!&=\!   \text{flops}\left(\mathbf{W}_a\right)\!+\!\text{flops}\left(\mathbf{u}_b\right)\!+\!\sum\nolimits_{k=1}^K\text{flops}\left(\mathbf{U}_k\right)\nonumber\\
  &\le\! \mathcal{O}\left(T^3\!+\!T^2KS\!+\!TKS^2\!+\!KS^3\right).\tag{29}
\end{align}

\subsection*{A.2  Feasibility Analysis of LM based IA Problem}
Recall that under assumption (11), and with the help of Lemma 3.1, the feasibility of the traditional LM based IA problem is equivalent to the solvability of polynomial equation (6),
corresponding to the formulated optimization model (22).

As discussed in Remark 3.5, Algorithm 1 can be used to check the sovability of (6).
If the objective value of (22) is zero, then (6) is feasible,
and if the objective value of (22) is greater than zero, then (6) is infeasible almost surely.

In addition to running the simulation of Algorithm 1,
we can also analyze the feasibility of the traditional LM based IA by counting the number of equations and variables involved in all subsets of equation (6).
Specifically, the definition of \emph{properness} is introduced in \cite{Feasibility_1}.
If the number of variables is greater than or equal to the number of equations in any subset of (6),
then (6) is classified as \emph{proper}, else as \emph{improper}.

From \cite{Feasibility_1, Feasibility_2, Feasibility_3}, we can conclude that if (6) is solvable, then it is proper, and conversely, if (6) is improper,
then it is infeasible.
In other words, properness is necessary but not sufficient for (6) to be solvable.
However, the characterization of properness may be computationally cumbersome since it requires to test all subsets of equations.
According to \cite{Feasibility_1},
the total number of equations involved in (6) can be calculated as
\begin{align}
N_E\!=\! \left(1\!+\!\sum\nolimits_{k=1}^K{d_k}\right)d_a.\tag{30}
\end{align}
Then there exist $2^{N_E}\!-\!1$ subsets of equations.
Testing each of them could be challenging due to its combinatorial nature.
For this reason, we will focus on the discussion of the necessary condition instead of proper condition for the feasibility of (6).

To simplify analysis, we define the following notations:
\begin{align}
  s_{M_a}&\!\triangleq\!\left(M_a\!-\!d_a\right)d_a,\tag{31}\\
  s_N&\!\triangleq\!\sum\nolimits_{k=1}^K\left(N_k\!-\!d_k\right)d_k,\, s_d \!\triangleq\! \sum\nolimits_{k=1}^{K}d_k.\tag{32}
\end{align}
Then the number of variables involved in (6) is given by
\begin{align}
N_V\!=\! N_b\!-\!1\!+\!s_{M_a} \!+\! s_N. \tag{33}
\end{align}
We summarize the above discussion in the following theorem.

\emph{Theorem 3.1 (necessary condition for the feasibility of the traditional LM based IA):}
Suppose that (11) is satisfied, and assume that the LM based IA problem is feasible (or equivalently, equation (6) is solvable).
Then $N_V\!\ge\!N_E$ holds, which is equivalent to the following condition:
\begin{align}
d_a^2\!-\!d_a\left(M_a\!-\!1\!-\!s_d\right) \!-\!\left(N_b\!-\!1\!+\!s_N \right)\!\le\!0. \tag{34}
\end{align}

\emph{Remark 3.6:}
The guiding insight of Theorem 3.1 is that, if necessary condition (34) does not hold, then the LM based IA is infeasible.
Therefore, the infeasibility of alignment can be determined without running the numerical simulation.

\emph{Remark 3.7:}
In the system model described in Section II,
we assume that the dimension of AN satisfies $1\!\le\!d_a\!\le\!M_a\!-\!1$.
By Theorem 3.1,
the selection of $d_a$ can be further refined as follows.
Note that (34) can be viewed as a quadratic constraint on $d_a$.
Then the corresponding discriminant is given by
\begin{align}
\Delta\!=\!\left(M_a\!-\!1\!-\!s_d\right)^2\!+\!4\left(N_b\!-\!1\!+\!s_N\right)\!>\!0. \tag{35}
\end{align}
Thus, to satisfy (34), the following relationship must hold.
\begin{align}
\left\lceil \frac{M_a\!-\!1\!-\!s_d\!-\!\sqrt{\Delta}}{2}\right\rceil\!\le\!d_a\!\le\!\left\lfloor \frac{M_a\!-\!1\!-\!s_d\!+\!\sqrt{\Delta}}{2}\right\rfloor. \tag{36}
\end{align}
From (35), we have that $\Delta\!>\!\left(M_a\!-\!1\!-\!s_d\right)^2$, it follows that
$M_a\!-\!1\!-\!s_d\!-\!\sqrt{\Delta}\!<\!0$.
Combining this with (36) and using $1\!\le\!d_a\!\le\!M_a\!-\!1$, we obtain the following condition
\begin{align}
1\!\le\!d_a\!\le\!\min\left(M_a\!-\!1,\left\lfloor \left(M_a\!-\!1\!-\!s_d\!+\!\sqrt{\Delta}\right)/{2}\right\rfloor\right). \tag{37}
\end{align}
It is also a necessary condition for the LM based IA.

\emph{Remark 3.8:} 
Suppose that (11) is satisfied, and assume that the LM based IA
is feasible (or equivalently, (6) is solvable).
Then by Theorem 3.1 and Remark 3.7, condition (37) is satisfied.
Equivalently, the LM based IA is infeasible if
\begin{align}
d_a\!>\!\min\left(M_a\!-\!1,\left\lfloor \left(M_a\!-\!1\!-\!s_d\!+\!\sqrt{\Delta}\right)/{2}\right\rfloor\right). \tag{38}
\end{align}
This condition can also be used to directly determine the infeasibility of alignment without running the numerical simulation.

\subsection*{A.3  Confidential Signal Cancellation Analysis}
Under the assumption that $M_a \!\ge\! 1\!+\! s_d$, we will theoretically  prove that the confidential signal cancellation phenomenon may occur.
The following theorem establishes a condition under which this phenomenon is inevitable almost surely.

\emph{Theorem 3.2:}
Consider the traditional LM based IA problem (6)--(8), and suppose that it is feasible (or equivalently, (6)--(8) are solvable).
If $d_a\!\ge\!M_a\!-\!s_d$, then the confidential signal cancellation will occur, i.e., $\mathbf{u}_b^H\mathbf{H}_{ba}\mathbf{v}_a\!=\!0$.
\begin{IEEEproof}
Suppose that $\mathbf{W}_a$, $\mathbf{u}_b$, and $\mathbf{U}_k$, $k\!=\!1,\ldots,K$, satisfy (6)--(8).
Thus we have that $\mathbf{M}\mathbf{W}_a\!=\!\mathbf{0}$ and $\bar{\mathbf{M}}\mathbf{v}_a\!=\!\mathbf{0}$,
where $\mathbf{M}$ and $\bar{\mathbf{M}}$ are defined in (12) and (13), respectively.

Note that $\mathbf{M}$ is a
square or fat
matrix of size $\left(1\!+\!s_d\right)\!\times\!M_a$.
Since $\text{span}\left(\mathbf{M}^H\right)\!\subseteq \!\mathbb{C}^{M_a}$ and since $\text{rank}\left(\mathbf{W}_a\right)\!=\!d_a$,
it follows by (6) that $\text{rank}\left(\mathbf{M}\right)\!=\!\text{dim}\left[\text{span}\left(\mathbf{M}^H\right)\right]\!\le\!M_a\!-d_a$,
and thus by assumption $d_a\!\ge\!M_a\!-\!s_d$, we have that $\text{rank}\left(\mathbf{M}\right)\!\le\!s_d$.
Hence, we conclude that rows of $\mathbf{M}$ are linearly dependent.

Also note that all channel matrices in $\mathbf{M}$ are assumed to be generic without any specialized structure.
Thus, any row of $\mathbf{M}$ can be expressed as a linear combination of other rows almost surely.
Note that $\mathbf{M}$ could be rewritten as $\mathbf{M}\!=\!\left[\mathbf{H}_{ba}^H\mathbf{u}_b, \bar{\mathbf{M}}^H\right]^H$, and hence
there must exist $\bm{\lambda}\!\in\!\mathbb{C}^{s_d}$ such that
$\mathbf{u}_b^H\mathbf{H}_{ba}\!=\!\bm{\lambda}^H\bar{\mathbf{M}}$.
Since $\bar{\mathbf{M}}\mathbf{v}_a\!=\!\mathbf{0}$,
it follows that
\begin{align}
  \mathbf{u}_b^H\mathbf{H}_{ba}\mathbf{v}_a\!=\!\left(\bm{\lambda}^H\bar{\mathbf{M}}\right)\mathbf{v}_a\!=\!\bm{\lambda}^H\left(\bar{\mathbf{M}}\mathbf{v}_a\right)\!=\!\mathbf{0},\tag{39}
\end{align}
which establishes the desired result.
\end{IEEEproof}

\emph{Remark 3.9:}
Assuming that all channels are generic without any special structure,
the traditional LM based IA takes for granted that rank conditions (9)--(10) are satisfied,
and thus only deals with IA equations (6)--(8).
However, Theorem 3.2 shows that under the assumption that (6)--(8) are satisfied,
if the dimension of AN is sufficiently large, then condition (9) does not hold, resulting in the confidential signal cancellation phenomenon.
This indicates the ineffectiveness of the traditional LM based IA for the AN assisted confidential transmission.
In addition to the theoretical analysis in Theorem 3.2, we also demonstrate the existence of this phenomenon in our numerical experiments in Section III-C.

\emph{Remark 3.10:} 
The confidential signal cancellation phenomenon can also be explained as follows.
First note that under the assumption that all channels are generic without any special structure, if the transmit and receive filters are designed without regard to the direct channel,
then the corresponding rank requirement is satisfied with probability $1$ (please see \cite[Section VII]{LM} for the detailed discussion and analysis).
Note that $\mathbf{H}_{kk}$ is the direct channel from Tx $k$ to Rx $k$, $k\!=\!1,\ldots,K$,
and since $\mathbf{V}_k$ and $\mathbf{U}_k$ in IA equations (6)--(8) are independent of $\mathbf{H}_{kk}$,
it follows by \cite{LM} that (10) is satisfied almost surely when performing the traditional LM based IA.
However, note that $\mathbf{H}_{ba}$ is the direct channel from Alice to Bob.
Since $\mathbf{W}_a$ and $\mathbf{u}_b$ in equation (6) are dependent of direct channel $\mathbf{H}_{ba}$,
it follows that rank condition (9) may not hold, and thus resulting in the confidential signal cancellation phenomenon.

\emph{Remark 3.11:} 
As discussed in Remark 3.10, it is reasonable and practical to assume that (10) holds almost surely when performing the LM based IA over generic channels.
Nevertheless, (9) needs to be taken into account in the transceiver design to circumvent the confidential signal cancellation.

\subsection{Analysis of the Modified IA}
From Remark 3.11, (10) holds almost surely, and thus the original IA problem (6)--(10) reduces to (6)--(8), subject to (9).
The imposition of (9) makes the LM based iterative algorithm difficult to develop directly.
However, to satisfy (9), it suffices to modify the LM based IA by implementing the MEB for the confidential transmission, thus preventing the confidential signal cancellation.
Then the original IA problem reduces to (6)--(8), where $\mathbf{u}_b$ ($\mathbf{v}_a$) is chosen to be the left (right) singular vector corresponding to the largest singular value of $\mathbf{H}_{ba}$.

For
any
$\mathbf{U}_k$, $k\!=\!1,\ldots,K$,
it follows by (11) and Lemma 3.1 that there always exists $\mathbf{V}_k$,
satisfying that ${\mathbf{M}_k\mathbf{V}_k}\!=\!\mathbf{0}$ and $\mathbf{V}_k^H\mathbf{V}_k\!=\!\mathbf{I}_{d_k}$, i.e., (8) holds trivially by ZF method.
Then the modified IA problem is equivalent to solve (6)--(7), and thus we can formulate the following optimization problem:
\begin{align}
  &\mathop{\min}_{\mathbf{W}_a, \mathbf{U}_k}
  \Vert{\mathbf{M}\mathbf{W}_a}\Vert_F^{2}\!+\!\Vert{\bar{\mathbf{M}}\mathbf{v}_a}\Vert^{2}\nonumber\\
 &\ \  \text{s.t.}\quad\ \mathbf{W}_a^H\mathbf{W}_a\!=\!\mathbf{I}_{d_a};\,\mathbf{U}_k^H\mathbf{U}_k\!=\!\mathbf{I}_{d_k},k\!=\!1,\ldots,K.\tag{40}
\end{align}

\emph{Remark 3.12:} 
Problem (40) can also be considered as a MIMO downlink interference mitigation problem.
The corresponding system model is shown in Fig. 2,
where a base station (Alice) transmits both the confidential signal and the AN signal.
The objective is to design $\mathbf{W}_a$ and $\mathbf{U}_k$, $k=1,\ldots,K$, to mitigate the AN and interference at all legitimate receivers.

By partitioning optimization variables into two blocks
($\mathbf{W}_a$ and $\left\{\mathbf{U}_k\right\}_{k=1}^K$),
alternating minimization \cite{NLP} can be developed
to facilitate the solution process.
In particular,
problem (40) is decoupled into the following two subproblems:
\begin{align}
  &\mathop{\min}_{\mathbf{W}_a}  \Vert{\mathbf{M}\mathbf{W}_a}\Vert_F^{2}
  \quad  \text{s.t.}\ \mathbf{W}_a^H\mathbf{W}_a\!=\!\mathbf{I}_{d_a}.\tag{41}  \\
  &\mathop{\min}_{\mathbf{U}_k}  \Vert{\bar{\mathbf{M}}\left[\mathbf{W}_a,\mathbf{v}_a\right]}\Vert_F^{2}\!=\!\sum\nolimits_{k=1}^K\Vert\mathbf{U}_k^H\mathbf{H}_{ka}\left[\mathbf{W}_a,\mathbf{v}_a\right]\Vert_F^2\nonumber\\
  &\ \text{s.t.}\ \ \mathbf{U}_k^H\mathbf{U}_k\!=\!\mathbf{I}_{d_k},k\!=\!1,\ldots,K.\tag{42}
\end{align}
Then we can solve (40) by solving (41) and (42) iteratively.
The execution process can be summarized in Algorithm 2.

\begin{algorithm}[h]\label{Algorithm 2}
\caption{Solving problem (40) under assumption (11)}
\begin{enumerate}
\item Initialization: Given local CSI at each legitimate transmitter and receiver.
\item Set $\mathbf{u}_b$ ($\mathbf{v}_a$) as the left (right) singular vector corresponding to the largest singular value of $\mathbf{H}_{ba}$.
Randomly generate unitary receive filters $\mathbf{U}_k$, $k\!=\!1\ldots,K$.
\item \textbf{Repeat}
    \begin{enumerate}
    \item Set columns of $\mathbf{W}_a$ as the $d_a$ least dominant eigenvectors of $\mathbf{M}^H\mathbf{M}$.
    \item Set columns of $\mathbf{U}_k$ as the $d_k$ least dominant eigenvectors of $\mathbf{H}_{ka}\left[\mathbf{W}_a,\mathbf{v}_a\right]\left[\mathbf{W}_a,\mathbf{v}_a\right]^H\mathbf{H}_{ka}^H$, $\forall k\!\in\!\mathcal{K}$.
    \end{enumerate}
\item \textbf{Until} the objective value of (40) converges or maximum number of iterations is reached.
\item
Set columns of $\mathbf{V}_k$ to be an orthonormal basis for $\text{null}\left(\mathbf{M}_k\right)$, $k\!=\!1,\ldots,K$.
\end{enumerate}
\end{algorithm}

Note that the objective value of (40) decreases with iteration 
and the objective function is lower bounded by zero, it follows that Algorithm 2 converges.
However, the global optimum cannot be guaranteed due to the nonconvexity of (40).

\emph{Remark 3.13:}
Recall that under (11) and using MEB for the confidential transmission, the original IA problem is reduced from (6)--(10) to (6)--(7).
To solve (6)--(7), we formulate an optimization model (40) and develop Algorithm 2.
However, the equivalence between (40) and (6)--(7) may not hold.
If the objective value of (40) is zero, then the solution obtained also satisfies equations (6)--(7).
Otherwise, (6)--(7) is infeasible almost surely due to the generic channel assumption.


As a distinct feature of Algorithm 2, it
not only provides the best confidential signal quality in terms of SNR, but also prevents the confidential signal cancellation, which is
not considered in the existing IA security approaches \cite{IA_PHY_Review, IA_CRN, IA_UAV, IA_AN_1, IA_AN_2, IA_AN_3}.

Note that only $\mathbf{W}_a$ and $\mathbf{U}_k$, $k\!=\!1,\ldots,K$, are involved in the iteration.
Since $\text{flops}\left(\mathbf{W}_a\right)$ is the same with (24), and $\text{flops}\left(\mathbf{U}_k\right)$ is
$126N_k^3\!+\!N_k^2\left(4d_a\!+\!3\right)\!+\!N_k\left(1\!+\!d_a\right)\left(8M_a\!-\!2\right)$,
then the number of flops for each iteration is given by
\begin{align}
\text{Flops}
\!&=\!   \text{flops}\left(\mathbf{W}_a\right)\!+\!\sum\nolimits_{k=1}^K\text{flops}\left(\mathbf{U}_k\right)\nonumber\\
  &\le\! \mathcal{O}\left(T^3\!+\!T^2KS\!+\!TKS^2\!+\!KS^3\right).\tag{43}
\end{align}


Note that the feasibility of the modified IA is equivalent to the solvability of (6)–(7).
Let E1 and E2 denote equations (6) and (7), respectively, and let E3 be equations combining E1 and E2.
Then we have three subsets of equations E1--E3.

If the modified IA is feasible,
the total number of variables in each subset is greater than or equal to the total number of equations.
Thus, we have the following three conditions:
\begin{align}
  &s_N\!+\!s_{M_a}\!\ge\!\left(1\!+\!s_d\right)d_a,\tag{44}\\
  &s_N\!\ge\!s_d,\tag{45}\\
  &s_N\!+\!s_{M_a}\!\ge\!\left(1\!+\!s_d\right)d_a\!+\!s_d.\tag{46}
\end{align}
Note that (45) holds since $1\!\le\!d_k\!\le\!N_k\!-\!1$, $\forall k\!\in\!\mathcal{K}$.
Since (46) implies (44), conditions (44)--(46) can be reduced to (46), which is equivalent to a quadratic constraint on $d_a$:
\begin{align}
d_a^2\!-\!d_a\left(M_a\!-\!1\!-\!s_d\right) \!-\!\left(s_N\!-\!s_d \right)\!\le\!0. \tag{47}
\end{align}
The above facts can be summarized in the following theorem.

\emph{Theorem 3.3 (necessary condition for the feasibility of the modified IA):}
Suppose that (11) is satisfied, and assume the modified IA is feasible (or equivalently, equations (6)--(7) are solvable).
Then condition (47) is satisfied.

\emph{Remark 3.14:}
The guiding insight of Theorem 3.3 is that, if (47) is not satisfied, then the modified IA is infeasible.
By Theorem 3.3, the selection of the dimension of AN can also be refined.
Note that (47) can be viewed as a quadratic constraint on $d_a$, and the corresponding discriminant is given by
\begin{align}
  \Delta\!=\!\left(M_a\!-\!1\!-\!s_d\right)^2\!+\!4\left(s_N\!-\!s_d\right)\!\ge\!0.\tag{48}
\end{align}
Since $1\!\le\!d_a\!\le\!M_a\!-\!1$, it follows by (47) that
\begin{align}
1\!\le\!d_a\!\le\!\min\left(M_a\!-\!1,\left\lfloor \left(M_a\!-\!1\!-\!s_d\!+\!\sqrt{\Delta}\right)/{2}\right\rfloor\right). \tag{49}
\end{align}

\emph{Remark 3.15:} 
Suppose that (11) is satisfied and the modified IA is feasible (or equivalently, (6)--(7) is solvable).
Then by Theorem 3.3,
the modified IA is infeasible if
\begin{align}
d_a\!>\!\min\left(M_a\!-\!1,\left\lfloor \left(M_a\!-\!1\!-\!s_d\!+\!\sqrt{\Delta}\right)/{2}\right\rfloor\right). \tag{50}
\end{align}
This condition can also be used to directly determine the infeasibility of alignment without running the numerical simulation.

\begin{figure}[t]
\centering
\includegraphics[height=6.11cm]{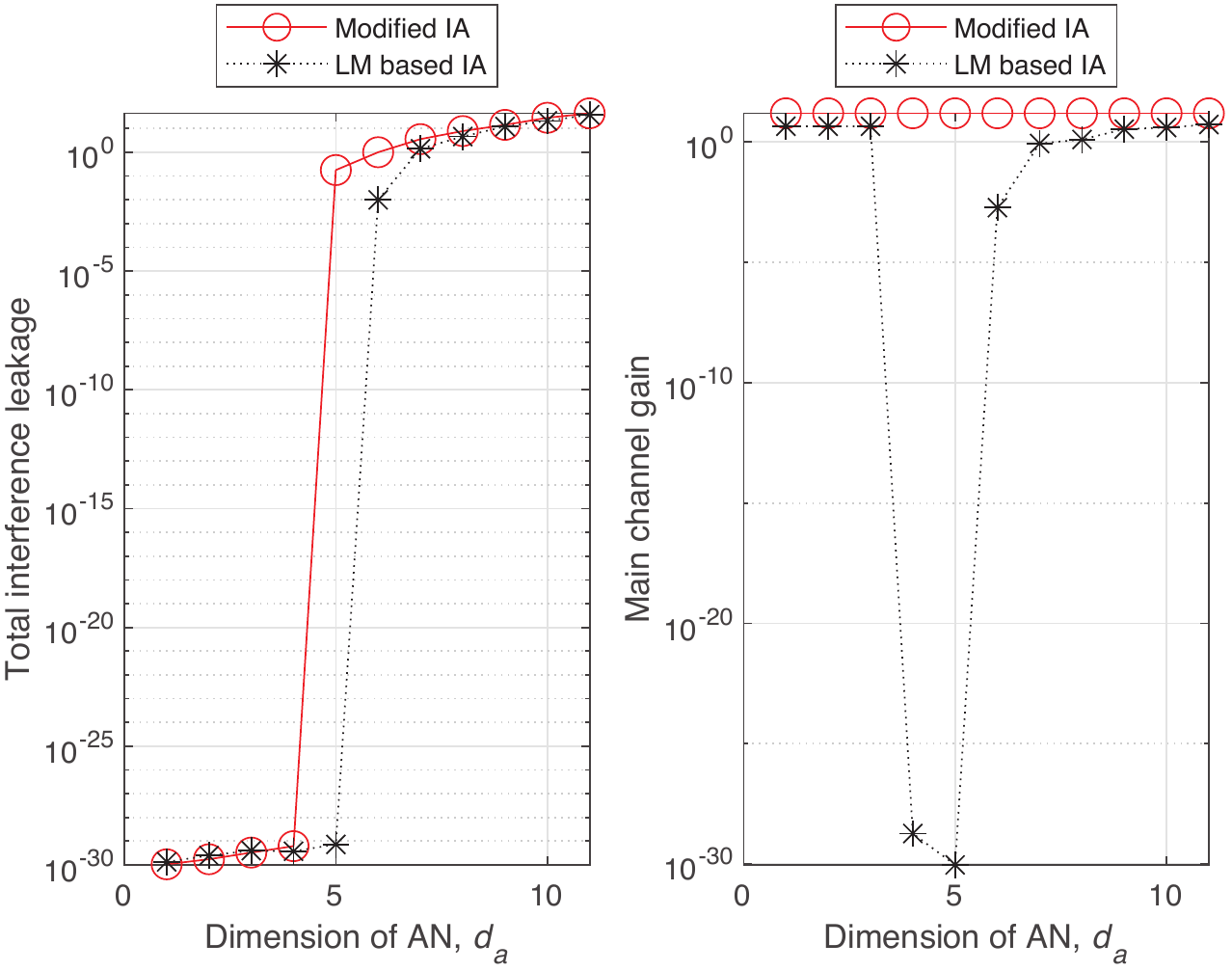}
\caption{Total interference leakage and main channel gain versus $d_a$,
with the network configuration $\mathcal{X}\!=\!\left(12\!\times\!2,\left[1,d_a\right]\right)\left(9\!\times\!4,2\right)^4$.}
\end{figure}

\begin{figure}[t]
\centering
\includegraphics[height=6.1cm]{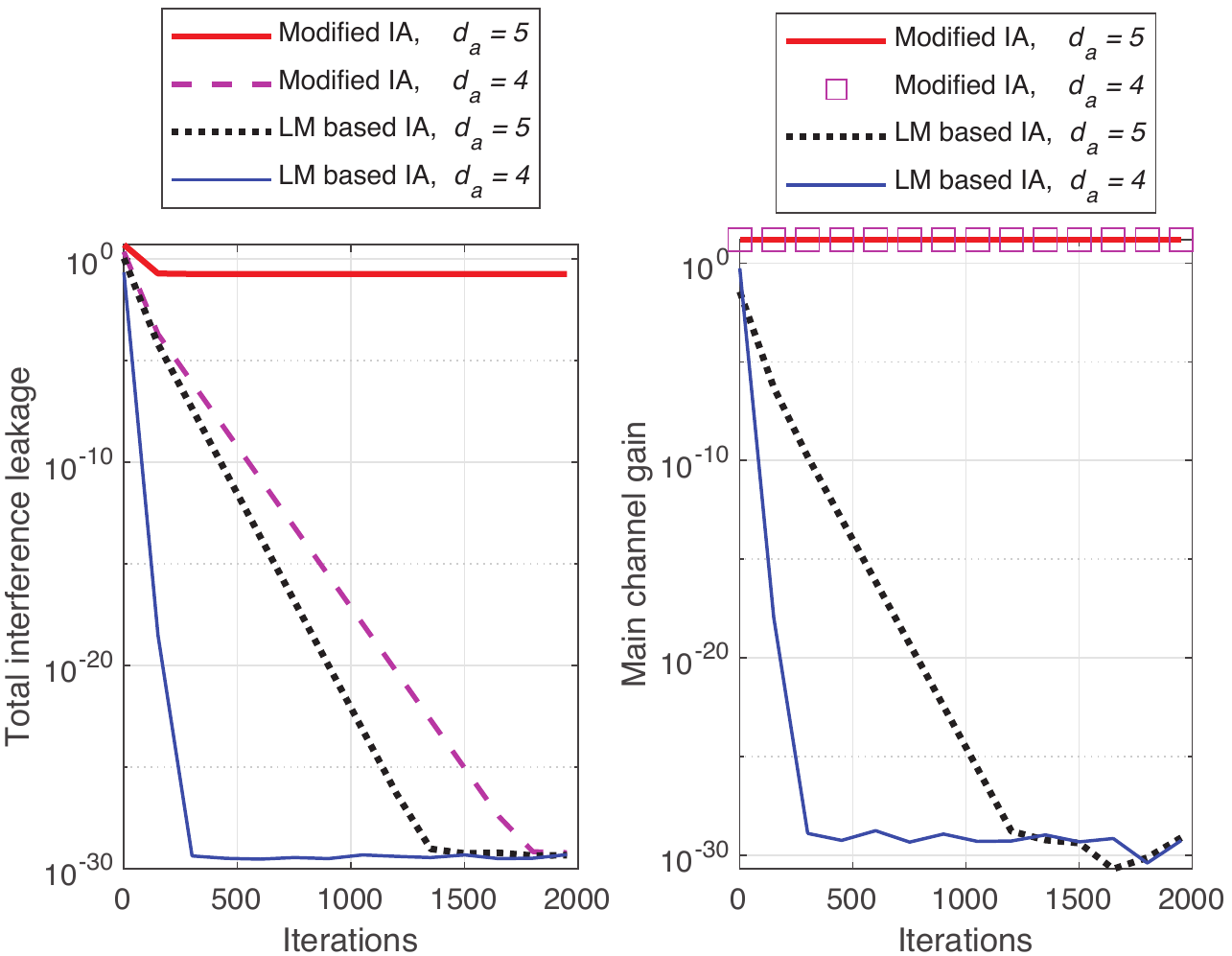}
\caption{Total interference leakage and main channel gain versus iterations,
with the network configuration $\mathcal{X}\!=\!\left(12\!\times\!2,\left[1,d_a\right]\right)\left(9\!\times\!4,2\right)^4$.}
\end{figure}

\subsection{Performance Analysis}
The performance measures are the total interference leakage and the main channel gain.
The total interference leakage is denoted by the objective values of problems (40) and (22) for the modified and traditional approaches, respectively,
and the main channel gain is denoted by $\vert\mathbf{u}_b^H\mathbf{H}_{ba}\mathbf{v}_a\vert^2$.
Unless otherwise specified, 
system parameters are set to be $K\!=\!4$, $M_a\!=\!12$, $N_b\!=\!2$, $M_k\!=\!9$, $N_k\!=\!4$, and $d_k\!=\!2$, for all $k\!\in\!\mathcal{K}$.
Note that in this setting, assumption (11) is satisfied.

Figs. 3 and 4 present numerical results of the total interference leakage and the main channel gain.
It can be observed that
for the modified scheme, IA is feasible when $1\!\le\!d_a\!\le\!4$,
and it is infeasible when $d_a\!>\!4$, which is consistent with Remark 3.15.
Moreover, the confidential signal cancellation can be prevented with the modified scheme.
For the traditional LM based method, IA is feasible when $1\!\le\!d_a\!\le\!5$, and it is infeasible when $d_a\!>\!5$, which is in agreement with Remark 3.8.
Although the main channel gain can be improved for the traditional LM based IA (as discussed in Remark 3.4),
the confidential signal cancellation cannot be avoided when $4\!\le\!d_a\!\le\!5$, which is consistent with Theorem 3.2.

\begin{figure}[t]
\centering
\includegraphics[height=6.19cm]{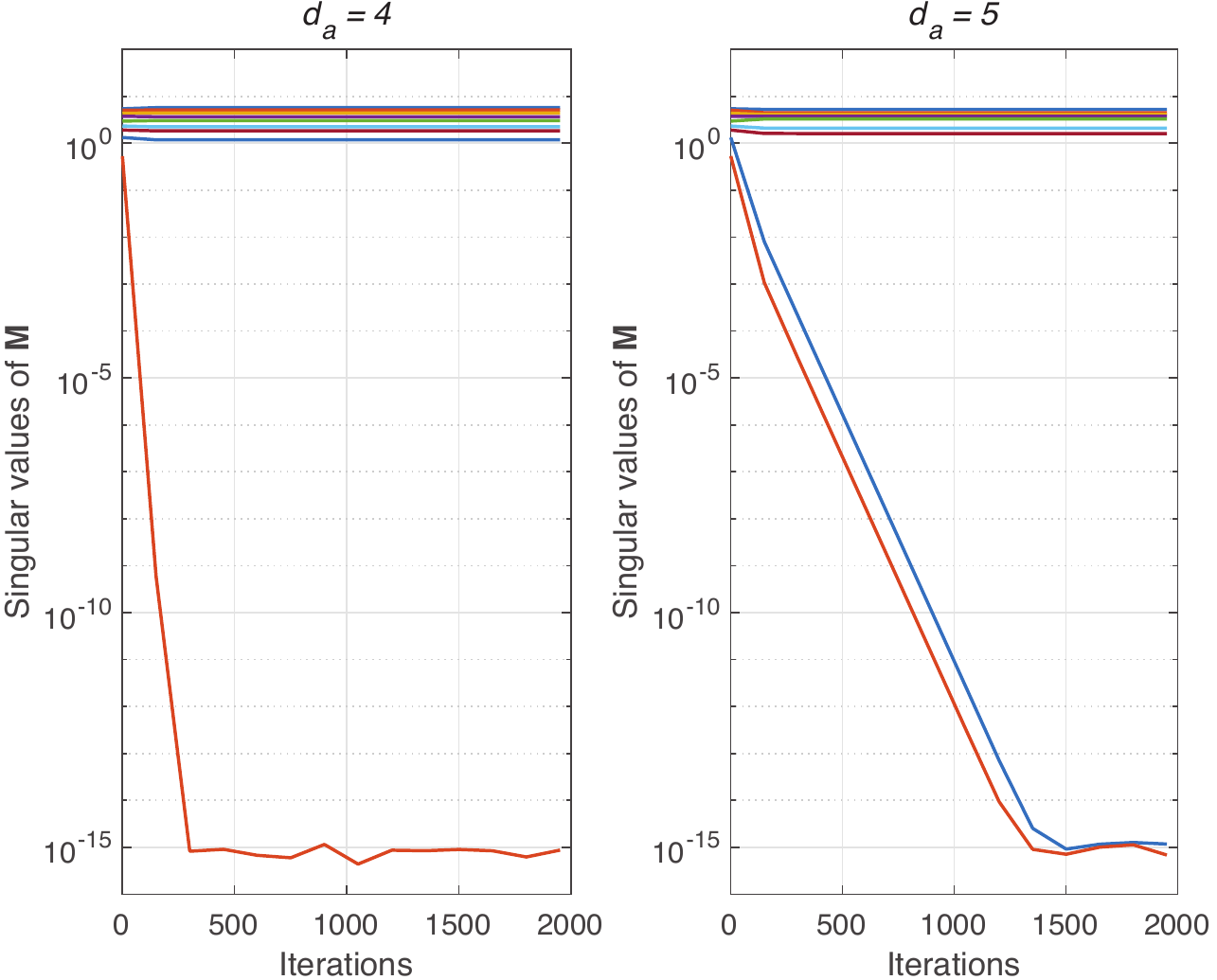}
\caption{ Singular values of $\mathbf{M}$ versus iterations for the traditional LM based IA,
with the network configuration $\mathcal{X}\!=\!\left(12\!\times\!2,\left[1,d_a\right]\right)\left(9\!\times\!4,2\right)^4$.  }
\end{figure}

\begin{figure}[t]
\centering
\includegraphics[height=6.14cm]{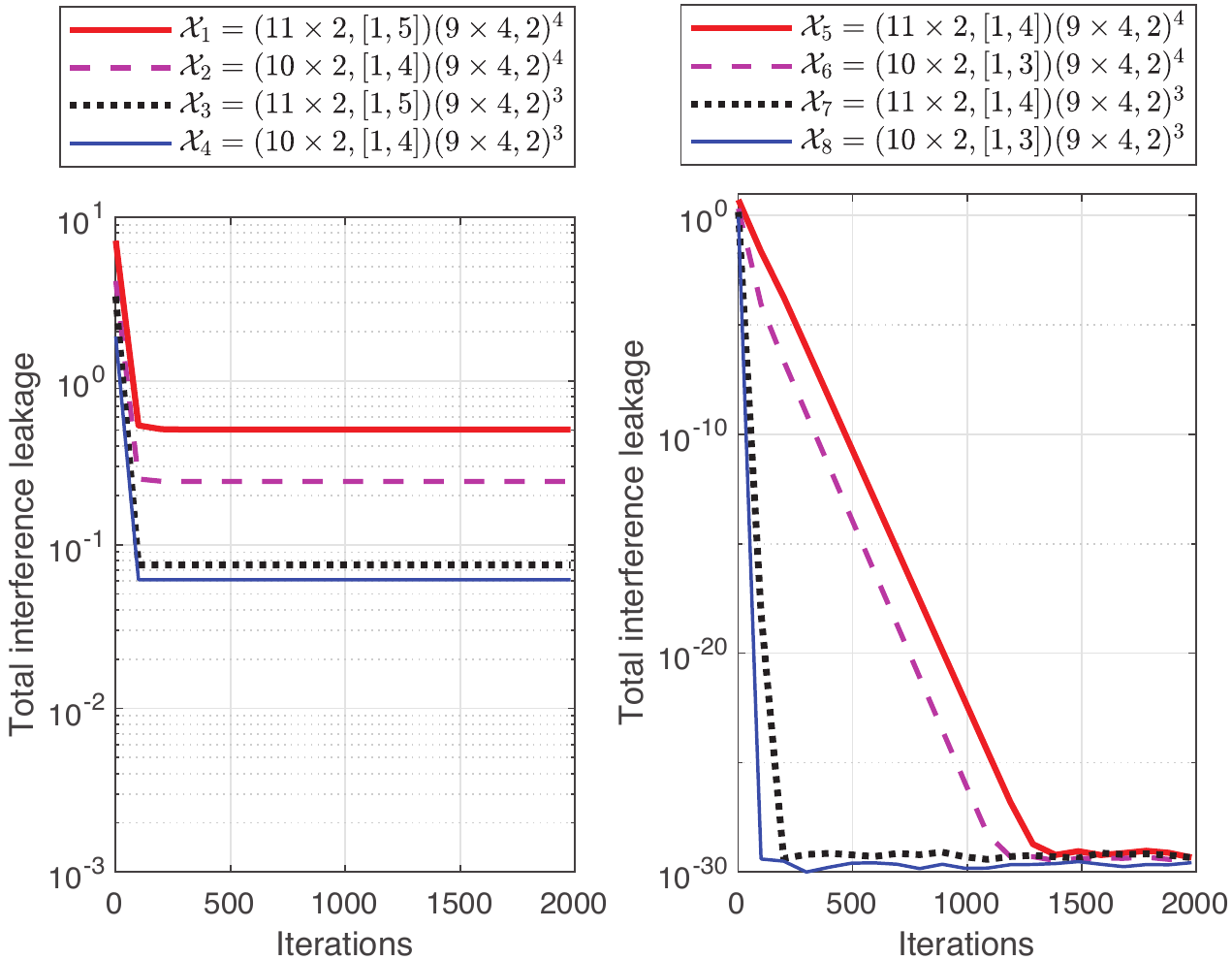}
\caption{Total interference leakage  versus iterations for the modified IA over different network configurations.}
\end{figure}

Fig. 5 shows
singular values of $\mathbf{M}$
for the traditional LM based IA.
Note that $\mathbf{M}$ becomes rank deficient with iterations,
and thus rows of $\mathbf{M}$ are dependent,
resulting in the confidential signal cancellation, as discussed in the proof of Theorem 3.2.
This phenomenon can also be observed in Figs. 3 and 4.


Fig. 6 provides the total interference leakage of the modified IA for different configurations.
It can be observed that
\begin{align}
&\text{when } d_a\!>\!4, \text{ } \left(11\!\times\!2,\left[1,d_a\right]\right)\left(9\!\times\!4,2\right)^4 \text{ is infeasible};   \nonumber\\
&\text{when } d_a\!>\!3, \text{ } \left(10\!\times\!2,\left[1,d_a\right]\right)\left(9\!\times\!4,2\right)^4 \text{ is infeasible};   \nonumber\\
&\text{when } d_a\!>\!4, \text{ } \left(11\!\times\!2,\left[1,d_a\right]\right)\left(9\!\times\!4,2\right)^3 \text{ is infeasible};   \nonumber\\
&\text{when } d_a\!>\!3, \text{ } \left(10\!\times\!2,\left[1,d_a\right]\right)\left(9\!\times\!4,2\right)^3 \text{ is infeasible}.   \nonumber
\end{align}
For the first two cases, numerical results are the same as theoretical results obtained with Remark 3.15.
However, for the last two cases, it follows by Remark 3.15 that
\begin{align}
&\text{when } d_a\!>\!5, \text{ } \left(11\!\times\!2,\left[1,d_a\right]\right)\left(9\!\times\!4,2\right)^3 \text{ is infeasible};   \nonumber\\
&\text{when } d_a\!>\!4, \text{ } \left(10\!\times\!2,\left[1,d_a\right]\right)\left(9\!\times\!4,2\right)^3 \text{ is infeasible}.   \nonumber
\end{align}
It shows that IA feasibility should be further studied to obtain a tighter condition, and we will explore this in future work.




\begin{figure}[t]
\centering
\includegraphics[height=6.19cm]{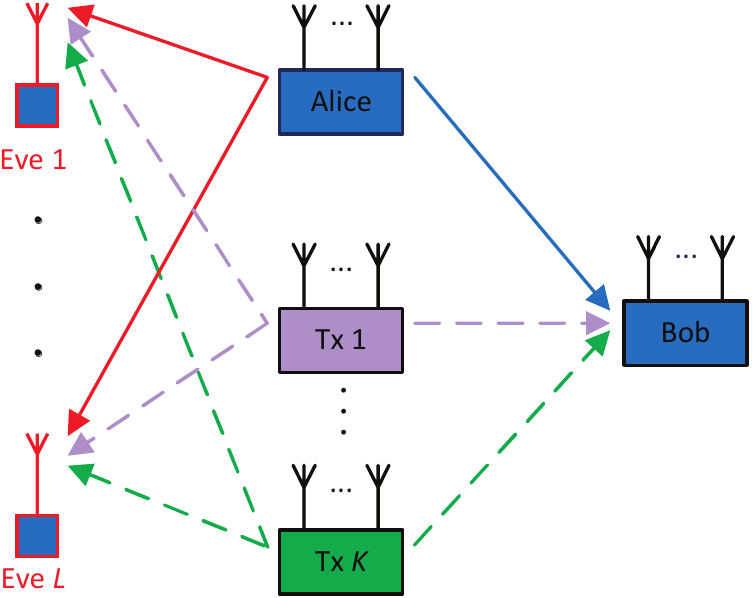}
\caption{The multi-user cooperative networks in the presence of non-colluding passive eavesdroppers.}
\end{figure}


\section{Design for Secrecy Rate Maximization}
The proposed modified IA described in Section III-B provides useful insights into the feasibility of alignment.
Under assumption (11), and suppose that the modified IA is feasible over the considered network configuration $\mathcal{X}$,
we will provide secrecy enhancing transmit design to maximize the secrecy rate performance, subject to a maximum allowable SOP.

The modified IA is assumed to be feasible, i.e., both AN and interfering signals at each legitimate receiver are eliminated.
Let $\sigma$ be the maximum singular value of $\mathbf{H}_{ba}$, and let $\sigma^2$ denote the main channel gain.
From (3) and (5), the SINRs at Bob and the $l$-th Eve, $l\!=\!1,\ldots,L$, are given by
\begin{align}
\gamma_b    \!&=\! \theta P_a\vert\mathbf{u}_b^H\mathbf{H}_{ba}\mathbf{v}_a\vert^2 \!=\! \theta P_a\sigma^2, \tag{51}\\
\gamma_{e_l}\!&=\! \frac{\theta P_a\vert\mathbf{h}_{e_l,a}^H\mathbf{v}_a\vert^2}
  {\frac{\left(1-\theta\right)P_a}{d_a}\Vert\mathbf{h}_{e_l,a}^H\mathbf{W}_a\Vert^2\!
  +\sum\nolimits_{k=1}^K\!\frac{P_k}{d_k}\Vert\mathbf{h}_{e_l,k}^H\mathbf{V}_k\Vert^2\!+\!1}.\tag{52}
\end{align}


Utilizing Wyner's wiretap code \cite{Wiretap_channel},
the confidential information transmitted by Alice is encoded before transmission.
Thus Alice determines the overall codeword rate $R_b$ and the secrecy rate $R_s$,
with the redundancy $R_e\!\triangleq\!R_b\!-\!R_s$ for anti-eavesdropping.
Note that as long as one Eve is able to wiretap the confidential message (i.e., $\log\left(1\!+\!\max_{l\in\mathcal{L}}\gamma_{e_l}\right)\!>\!R_e$),
then the information-theoretic secrecy cannot be guaranteed and secrecy outage occurs.
Hence, the SOP can be expressed as
\begin{align}
\varepsilon\!=\!\Pr\left\{\max\nolimits_{l\in\mathcal{L}}\gamma_{e_l}\!>\!2^{R_b\!-\!R_s}\!-\!1\right\}.  \tag{53}
\end{align}
Let $\varepsilon_{th}$ be a maximum allowable SOP. Then the SOP constrained SRM problem can be formulated as
\begin{align}
\max \ R_s \ \  \text{s.t.} \ \varepsilon\!\le\!\varepsilon_{th}; \ 0\!\le\!\theta\!\le\!1.\tag{54}
\end{align}

\emph{Remark 4.1:}
Note that problem (54) can also be considered as a cooperative SRM problem \cite{Cooperation_SRM}.
The corresponding system model is shown in Fig. 7, where Alice intends to transmit a confidential message to Bob,
in the presence of multiple Eves.
Different from \cite{Cooperation_SRM}, the transceiver filters are obtained by the modified IA,
which effectively mitigate both AN and multi-user interference while avoiding confidential signal cancellation.
Besides,
we consider a scenario with multiple cooperative transceivers, instead of a single cooperative jammer in \cite{Cooperation_SRM}.


Next,
we derive a closed form for the SOP constraint, and reformulate (54) as a power allocation problem.
Then, a numerical method is developed for the optimal solution.

\subsection{Reformulation of the SRM Problem (54)}
To simplify notations, new variables are defined as follows:
\begin{align}
  &\!\!\!\!Y_l\!\triangleq\!{\left(1\!-\!\theta\right)P_a}/{d_a}\Vert\mathbf{h}_{e_l,a}^H\!\mathbf{W}_a\Vert^2,
  Z_{l,k}\!\triangleq\!{P_k}/{d_k}\Vert\mathbf{h}_{e_l,k}^H\!\mathbf{V}_k\Vert^2, \tag{55}\\
  &\!\!\!\!X_l\!\triangleq\!\theta P_a\vert\mathbf{h}_{e_l,a}^H\mathbf{v}_a\vert^2,
  T_l\!\triangleq\! Y_l \!+\! \sum\nolimits_{k=1}^{K} Z_{l,k},
  l\!\in\!\mathcal{L}, k\!\in\!\mathcal{K}. \tag{56}
\end{align}
From \cite{BPT}, we have that
$Y_l\!\sim\!\Gamma\left(\alpha_a,\lambda_a\right)$,
$Z_{l,k}\!\sim\!\Gamma\left(\alpha_k,\lambda_k\right)$,
and $X_l\!\sim\!\text{Exp}\left(\lambda\right)$,
where
$\alpha_a\!\triangleq d_a$,
$\lambda_a\!\triangleq\!\alpha_a/\left(\left(1\!-\!\theta\right)P_a\right)$,
$\alpha_k\!\triangleq d_k$,
$\lambda_k\!\triangleq\!\alpha_k/P_k$,
and $\lambda\!\triangleq\!1/\left(\theta P_a\right)$.

By (52) and (56), we have $\gamma_{e_l}\!=\!X_l/\left(T_l\!+\!1\right)$, $l\!=\!1,\ldots,L$.
From Theorem 2 in \cite[Chapter 2.7]{BPT},
$\gamma_{e_1},\ldots,\gamma_{e_L}$ are independent, and thus
the SOP in (53) can be expressed as
\begin{align}
\!\varepsilon\!=\! 1 \!-\! \left(\Pr\left\{\gamma_{e_l}\!\le\!2^{R_b\!-\!R_s}\!-\!1\right\}\right)^L\!=\! 1 \!-\! \left(1 \!-\! \bar{F}_{\gamma_{e_l}} \left(\mu\right)\right)^L,   \tag{57}
\end{align}
where $\bar{F}_{\gamma_{e_l}}\left(r\right)$
denotes the CCDF
of $\gamma_{e_l}$ and $\mu\!\triangleq\!2^{R_b\!-\!R_{s}}\!-\!1$.

\emph{Theorem 4.1:} A closed form expression of (57) is given by
\begin{align}
  \!\!\!\varepsilon\!=\!1 \!-\! \left(1 \!-\! \frac{1}{e^{\lambda \mu}}\left(\frac{\lambda_a}{\lambda_a\!+\!\lambda \mu}\right)^{\alpha_a}
        \prod\nolimits_{k=1}^{K}\!\left(\frac{\lambda_k}{\lambda_k\!+\!\lambda \mu}\right)^{\alpha_k}\right)^L\!\!.       \tag{58}
\end{align}
\begin{IEEEproof}
Please refer to Appendix A.
\end{IEEEproof}

From (57),
$\varepsilon$ increases with $R_s$,
and thus the SOP constraint in (54) holds with equality at the maximum secrecy rate,
leading to $\varepsilon\!=\!\varepsilon_{th}$, and thus by (58) we have that
\begin{align}           
\!\!\!c
\!=\! \frac{w}{P_a}\!+\!\alpha_a\ln\left(\!1\!+\!\frac{\left(1\!-\!\theta\right)w}{\alpha_a}\!\right)\!+\!\sum\nolimits_{k=1}^{K}\alpha_k\ln\left(\!1\!+\!\frac{w}{g_k}\!\right),\tag{59}
\end{align}
where
$c\!\triangleq\!-\!\ln\left(1\!-\!(1\!-\!\varepsilon_{th})^{\frac{1}{L}}\right)$,
$w\!\triangleq\!\mu/\theta$, and $g_k\!\triangleq\!P_a\lambda_k$.

\begin{figure*}[hb]
\hrulefill
\begin{align}
&w'\left(\theta\right) \!=\! \frac{\alpha_a w\left(\theta\right)} {\alpha_a\!+\!\left(1\!-\!\theta\right)w\left(\theta\right)}
\left[
\frac{1}{P_a}
\!+\!
\frac{\left(1\!-\!\theta\right)\alpha_a} {\alpha_a\!+\!\left(1\!-\!\theta\right)w\left(\theta\right)}
\!+\!
\sum\nolimits_{k=1}^{K}\frac{\alpha_k}{g_k\!+\!w\left(\theta\right)}
\right]^{-1}. \tag{62}\\
&\frac{w'\left(\theta\right)} {w\left(\theta\right)} \!=\! {\alpha_a}
\left[
\frac{\alpha_a\!+\!\left(1\!-\!\theta\right)w\left(\theta\right)}{P_a}
\!+\!
\left(1\!-\!\theta\right)\alpha_a
\!+\!
\left[\alpha_a\!+\!\left(1\!-\!\theta\right)w\left(\theta\right)\right]\sum\nolimits_{k=1}^{K}\frac{\alpha_k}{g_k\!+\!w\left(\theta\right)}
\right]^{-1}. \tag{63}
\end{align}
\end{figure*}

Let $\gamma_B\!\triangleq\!P_a\sigma^2$. Then the objective of (54) is given by
\begin{align}           
R_s = R_b - \log(1+\mu) = \log\!\left(\frac{1+\theta\gamma_B}{1+\theta w}\right).\tag{60}
\end{align}
By (59), $w$ can be taken as a function of $\theta$, denoted by $w\left(\theta\right)$, and it can be verified that $w\left(\theta\right)\!>\!0$.
Then, $R_s$ in (60) can be regarded as a function of $\theta$, denoted by $R_s\left(\theta\right)$. Hence, (54) can be reformulated as the following power allocation problem
\begin{align}           
\max_{\theta} \ R_s(\theta) \!=\!\log\!\left(\!\frac{1\!+\!\theta\gamma_B}{1\!+\!\theta w(\theta)}\!\right)  \ \  \text{s.t.} \ (59); \ 0\!\le\!\theta\!\le\!1.\tag{61}
\end{align}

\subsection{Problem and Solution Analysis}

By analyzing constraint (59), we have the following results.

\emph{Lemma 4.1:} The function $w(\theta)$, determined by (59), is monotonically increasing with $\theta$.

\begin{IEEEproof}
From (59),
we can derive $w'(\theta)$ as in (62). It is positive, and thus establishing the desired result.
\end{IEEEproof}

\emph{Corollary 4.1:} The function $(1\!-\!\theta)w(\theta)$ determined by (59) is monotonically decreasing with $\theta$.

\begin{IEEEproof}
By Lemma 4.1, the first and third terms on the right-hand side of (59) increase with $\theta$.
To satisfy constraint (59), we have that $(1\!-\!\theta)w(\theta)$ decreases with $\theta$.
\end{IEEEproof}
\emph{Corollary 4.2:} The function ${w'(\theta)}/{w(\theta)}$ is monotonically increasing with $\theta$.
\begin{IEEEproof}
From (62), ${w'(\theta)}/{w(\theta)}$ is given by (63),
it follows by Lemma 4.1 and Corollary 4.1 that the denominator of (63) strictly decreases with $\theta$, establishing the result.
\end{IEEEproof}

\emph{Theorem 4.2:}
The objective of (61) is strictly concave on $\theta$.

\begin{IEEEproof}
Please refer to Appendix B.
\end{IEEEproof}

\emph{Corollary 4.3:} $R_s(\theta)$ in (60) is monotonically increasing when $R_{s}'(1)\!\ge\!0$, and decreasing when $R_{s}'(0^+)\!\le\!0$.
\begin{IEEEproof}
From Theorem 4.2,
it follows that
when $R_{s}'(1)\!\ge\!0$, $R_{s}'(\theta)\!>\!R_{s}'(1)\!\ge\!0$ for all $\theta\!\in\!(0,1)$, i.e., $R_{s}(\theta)$ increases with $\theta$.
In addition, when $R_{s}'(0^+)\!\le\!0$, we have $R_{s}'(\theta)\!<\!R_{s}'(0^+)\!\le\!0$ for all $\theta\!\in\!(0,1]$, meaning that $R_{s}(\theta)$ decreases with $\theta$.
\end{IEEEproof}

\emph{Remark 4.2:} $R_s\!>\!0$ is achievable only when $R_s'(0^+)\!>\!0$, which is equivalent to condition $\gamma_B\!>\!w(0^+)$.
Note that $w\left(0^+\right)$ can be obtained by solving equation (59).
In other words, a positive secrecy rate can be guaranteed only when $\gamma_B$ is larger than a threshold associated with SOP constraint (59).

%
\begin{IEEEproof}
Please refer to Appendix C.
\end{IEEEproof}

With the above results, the process of solving SRM problem (61) can be summarized in Algorithm 3.
%
%
\begin{algorithm}[h]\label{Algorithm 3}
\caption{Solving SRM Problem (61)}
\begin{enumerate}
\item Initialization: Given $\varepsilon_{th}$, transceiver strategies of legitimate users (which are obtained by Algorithm 2),
                      the CSI of $\mathbf{H}_{ba}$,
                      and the statistical CSIs of $\mathbf{h}_{e_l,a}$ and $\mathbf{h}_{e_l,k}$,
                      where $k\!\in\!\mathcal{K}$ and $l\!\in\!\mathcal{L}$.
\item Calculate $R_s'\left(1\right)$ according to (B1) in Appendix B.
\begin{itemize}
  \item If $R_s'\left(1\right)\!\ge\!0$, set $\theta^*\!=\!1$.
  \item Otherwise, calculate $R_s'\left(0^+\right)$ according to (B1).
  \begin{itemize}
    \item If $R_s'\left(0^+\right)\!\le\!0$, set $\theta^*\!=\!0$.
    \item Otherwise, solve equations $R_s'\left(\theta^*\right)\!=\!0$ and (59).
  \end{itemize}
\end{itemize}
\item The optimal power allocation ratio is determined as the result of 2),
      and then calculate $R_s\left(\theta^*\right)$ by (60).
\end{enumerate}
\end{algorithm}

Note that it is difficult to find an analytical expression for the optimal power allocation ratio $\theta^*$.
However, we can obtain an accurate approximation of $\theta^*$ in the high SNR region (i.e., $P_a$ is large), as stated in the following result.


\emph{Remark 4.3:}
When $P_a\!\rightarrow\!\infty$, the optimal solution to problem (61), denoted by $\theta^*$, can be approximated as in (64).
\begin{align}           
\theta^* \!\approx\! \left( 1\!+\!\sqrt{   {\alpha_a}{\left[1\!-\!(1\!-\!\varepsilon_{th})^{\frac{1}{L}}\right]^{-\!\frac{1}{\alpha_a}}} \!-\! \alpha_a   }\right)^{\!-\!1}.\tag{64}
\end{align}\par

\begin{IEEEproof}
Please refer to Appendix D.
\end{IEEEproof}


\subsection{Performance Analysis}

\begin{figure}[t]
\centering
\includegraphics[height=6.23cm]{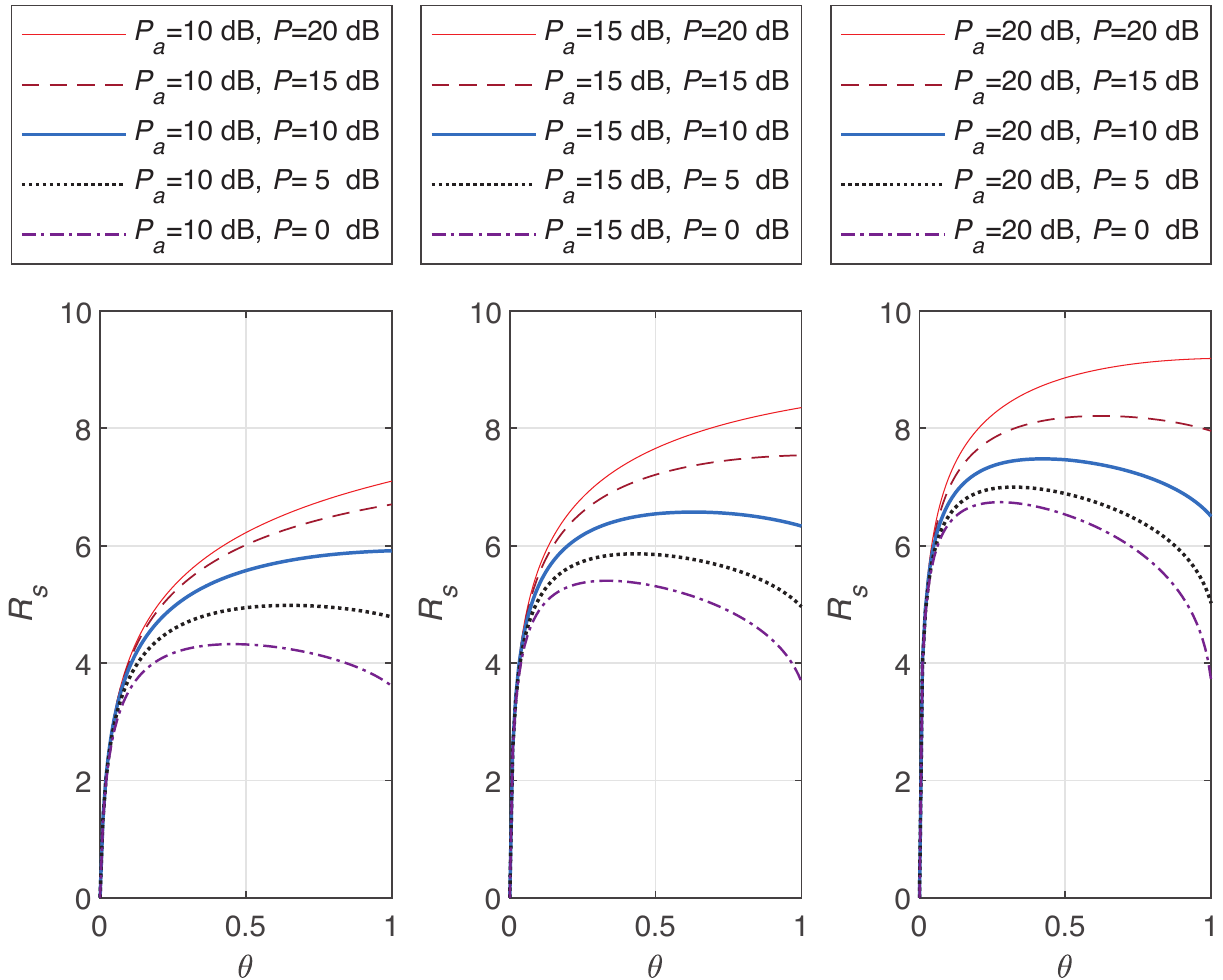}
\caption{Secrecy rate $R_s$ versus power allocation ratio $\theta$.}
\end{figure}

\begin{figure}[t]
\centering
\includegraphics[height=6.23cm]{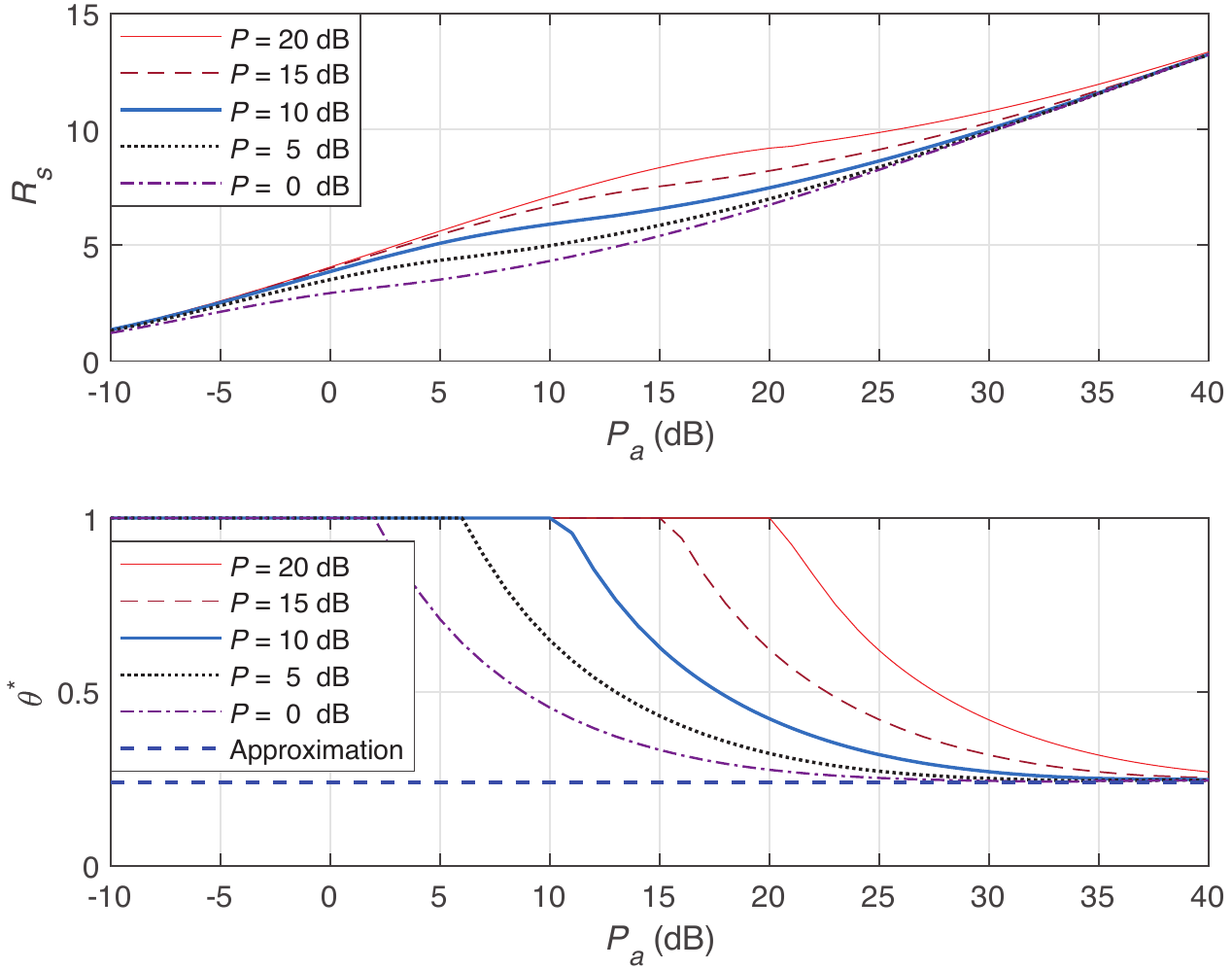}
\caption{Secrecy rate $R_s$ and optimal power allocation ratio $\theta^*$ versus $P_a$.}
\end{figure}

\begin{figure}[t]
\centering
\includegraphics[height=6.23cm]{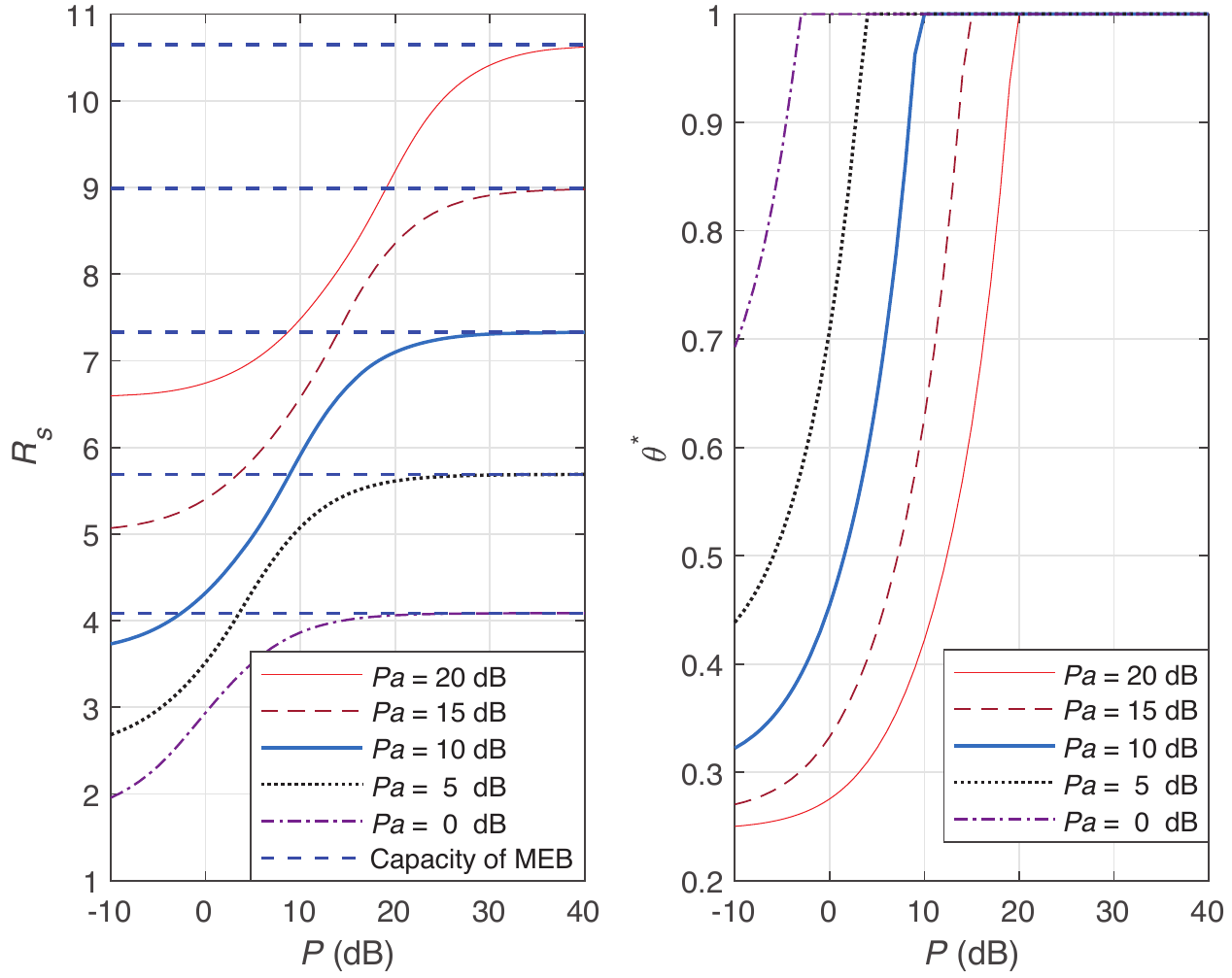}
\caption{Secrecy rate $R_s$ and optimal power allocation ratio $\theta^*$ versus $P$.}
\end{figure}

\begin{figure}[t]
\centering
\includegraphics[height=6.23cm]{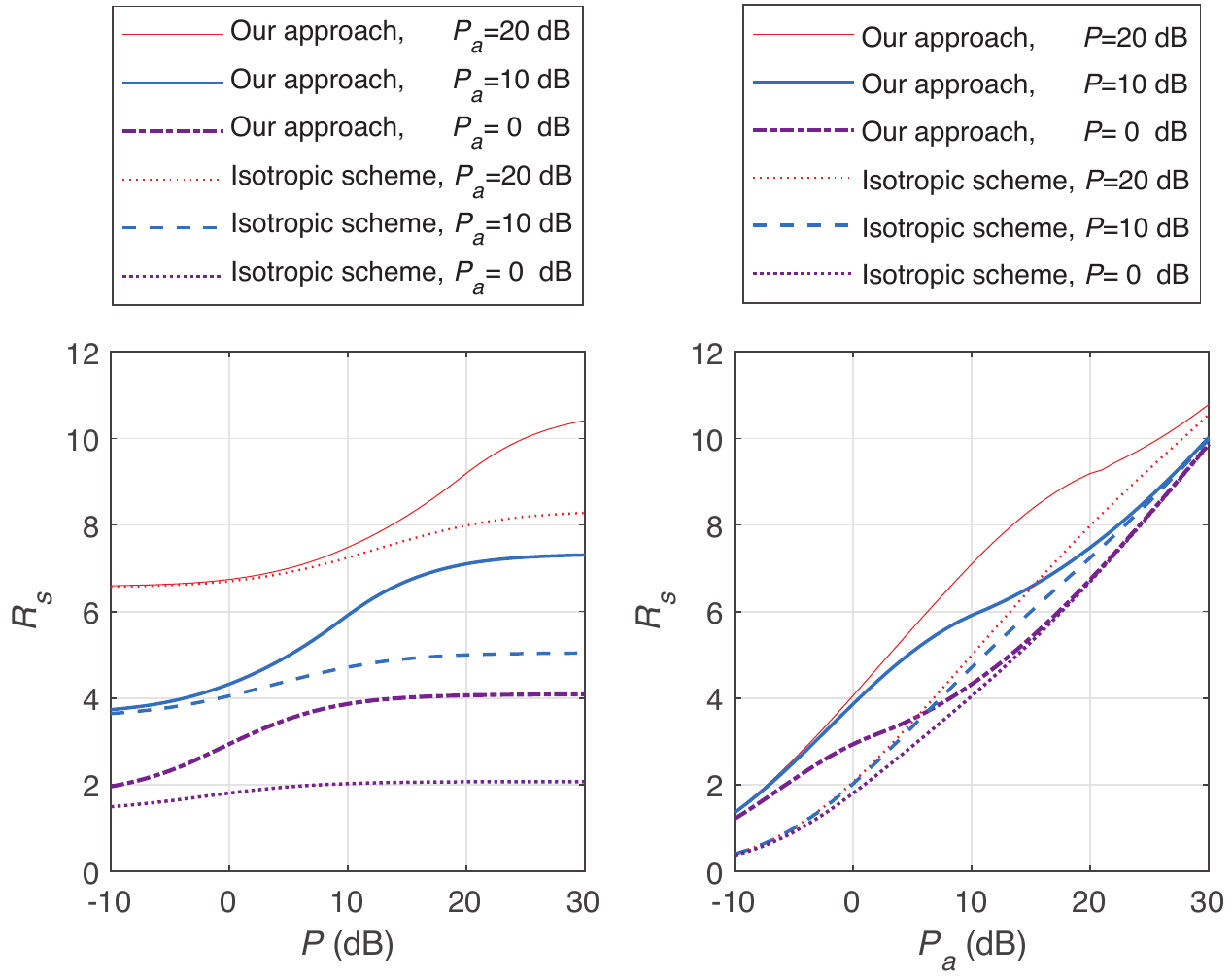}
\caption{Comparison of our approach with isotropic transmission scheme.}
\end{figure}

In this subsection, we demonstrate the secrecy rate performance of the proposed modified IA.
Network configuration is  $\mathcal{X}\!=\!\left(12\!\times\!2,\left[1,4\right]\right)\left(9\!\times\!4,2\right)^4$.
In this setting, assumption (11) is satisfied and the modified IA is feasible (see Sections III-B and III-C for details).
We set $L\!=\!16$, $\varepsilon_{th}\!=\!0.1$ and $P_k\!=\!P$, $\forall k\!\in\!\mathcal{K}$.
The main channel gain is set to be $\sigma^2\!=\!16$,
and the secrecy rate $R_s$ is measured by bits per channel use (bpcu).

Fig. 8 presents the impact of power allocation ratio between the confidential signal and the AN signal on the secrecy rate performance.
It can be seen that the secrecy rate is a strictly concave function of $\theta$, which is consistent with Theorem 4.2.

Fig. 9 shows secrecy rate performance and the corresponding optimal power allocation ratio.
Note that
$\theta^*\!=\!1$ means that full power of Alice is used for confidential signal.
In this case, the interference created by other legitimate users is strong enough to mask the transmission of the confidential message.
As $P_a$ keeps increasing, the probability of information leakage increases, and thus it is preferable to perform AN-assisted secrecy beamforming at Alice.
In addition, when $P_a$ becomes large, $\theta^*$ converges to a constant value, as discussed in Remark 4.3.
It can also be shown that the approximation is accurate.

Fig. 10 provides secrecy rate and the corresponding optimal power allocation ratio versus $P$.
Note that the capacity of MEB is the upper bound of secrecy rate when $P$ becomes large.

Fig. 11 compares our approach with isotropic transmission which is a heuristic scheme with transmit
covariance matrix denoted by $P_a/\left(1\!+\!d_a\right)\mathbf{I}_{1\!+d_a}$, i.e., $\theta\!=\!1/\left(1\!+\!d_a\right)$.
It can be shown that
the secrecy rate performance with our scheme outperforms that with isotropic scheme for a wide SNR region.


\section{Conclusion}
We studied physical layer security (PLS) in multi-user interference alignment (IA) networks.
We analyzed the cause of the confidential signal cancellation with the traditional leakage minimization (LM) based IA,
and then established a condition under which this phenomenon is inevitable.
Moreover, we proposed a cure to overcome this obstacle, by integrating the max-eigenmode beamforming (MEB) into the LM based IA.
The necessary condition for IA feasibility was proposed and useful guidelines for the selection of the AN dimension was also provided.
Furthermore, the explicit transmit design for the secrecy rate maximization (SRM) was investigated.
Numerical results confirmed the effectiveness of the proposed approach.

\appendices
\section{Proof of Theorem 4.1}       
\begin{IEEEproof}
Recall that $\gamma_{e_l}\!=\!X_l/\left(T_l\!+\!1\right)$,
where $X_l\!\sim\!\text{E}\left(\lambda\right)$,
$T_l\!=\! Y_l \!+\! \sum\nolimits_{k=1}^{K} Z_{l,k}$.
where $Y_l\!\sim\!\Gamma\left(\alpha_a,\lambda_a\right)$, $Z_{l,k}\!\sim\!\Gamma\left(\alpha_k,\lambda_k\right)$,
for all $l\!\in\!\mathcal{L}$, $k\!\in\!\mathcal{K}$.
Let $f_{T_l}$ and $F_{T_l}$ be the probability density function (PDF)
and the cumulative distribution function (CDF) of $T_l$, respectively,
and let $f_{X_l}$ represent the PDF of $X_l$.
Then the CCDF of $\gamma_{e_l}$, denoted by $\bar{F}_{\gamma_{e,l}}\left(r\right)$, can be calculated as
\begin{align}
\bar{F}_{\gamma_{e,l}}\left(r\right)
\!&=\!  \Pr\left\{ \frac{ X_l}{T_l\!+\!1}\!>\!r \right\}\nonumber\\
\!&=\!  \Pr\left\{T_l\!<\!X_l/r\!-\!1\right\}   \nonumber\\
  &=\!  \int_{r}^{\infty} {   \int_{0}^{x/r-1}{f_{T_l}\left(t\right)}dt\,f_{X_l}\left(x\right)}dt     \nonumber\\
  &=\!  \frac{\lambda r}{e^{\lambda r}}\int_{r}^{\infty}
        { F_{T_l}\left(\frac{x}{r}\!-\!1\right)  e^{-\lambda r \left(\frac{x}{r}\!-\!1\right)} }d\left(\frac{x}{r}\!-\!1\right)\nonumber\\
  &=\!  \frac{\lambda r}{e^{\lambda r}}\int_{0}^{\infty}
        {F_{T_l}\left(t\right)} e^{-\left(\lambda r\right) t} dt,  \tag{A1}
\end{align}
where the third equality follows from the fact that $T_l$ and $X_l$ are nonnegative random variables,
and the fourth equality follows from the fact that the CDF is the integral of the PDF.

Note that $F_{T_l}$ is the integral of $f_{T_l}$,
it follows by the definition of the Laplace transform \cite[12.11]{Table} that
\begin{align}
\!      \int_{0}^{\infty}   {F_{T_l}\left(t\right)} e^{-\left(\lambda r\right) t} dt
\!=\!  L\left[\int_{0}^{t}f_{T_l}\left(x\right)dx; s\right],\tag{A2}
\end{align}
where $s\!=\!\lambda r$. According to the property given in \cite[12.12]{Table},
we can obtain the following expression
\begin{align}
L\left[\int_{0}^{\infty}f_{T_l}\left(t\right)dt; s\right]
\!&=\!  \frac{1}{s} L\left[f_{T_l}\left(t\right); s\right]\nonumber\\
\!&=\!  \frac{1}{s} \int_{0}^{\infty} {f_{T_l}\left(t\right) e^{-st}dt}\nonumber\\
\!&=\!\frac{1}{s}\mathbb{E}\left(e^{-sT_l}\right),\tag{A3}
\end{align}
where
$\mathbb{E}\left(\cdot\right)$ represents the expectation operation.
Note that $Y_l$, $Z_{l,1},\ldots,Z_{l,K}$ are independent gamma random variables,
and we have the following
\begin{align}
\mathbb{E}\left(e^{-sT_l}\right)
&\!=\!\mathbb{E}\left(e^{-s\left(Y_l \!+\! \sum\nolimits_{k=1}^{K} Z_{l,k}\right)}\right)\nonumber\\
&\!=\!\mathbb{E}\left(e^{-sY_l}\right)\!\prod\nolimits_{k=1}^{K}\!\mathbb{E}\left(e^{-sZ_{l,k}}\right)\nonumber\\
&\!=\! L\left[f_{Y_l}\left(t\right); s\right]\!\prod\nolimits_{k=1}^{K}\!L\left[f_{Z_{l,k}}\left(t\right); s\right]\nonumber\\
&\!=\!\left(\frac{\lambda_a}{\lambda_a\!+\!s}\right)^{\alpha_a}
        \prod\nolimits_{k=1}^{K}\!\left(\frac{\lambda_k}{\lambda_k\!+\!s}\right)^{\alpha_k},\tag{A4}
\end{align}
where the last equality follows from the Laplace transform of the gamma PDF.
From (A1)--(A4), we have that
\begin{align}
        \bar{F}_{\gamma_{e,l}}\left(r\right)
  \!=\!  \frac{1}{e^{\lambda r}}\left(\frac{\lambda_a}{\lambda_a\!+\!\lambda r}\right)^{\alpha_a}
        \prod\nolimits_{k=1}^{K}\!\left(\frac{\lambda_k}{\lambda_k\!+\!\lambda r}\right)^{\alpha_k}.\tag{A5}
\end{align}
Substituting (A5) into (57), we obtain that
\begin{align}
\!\!\!\!\varepsilon\!&=\! 1 \!-\! \left(\!1 \!-\! \bar{F}_{\gamma_{e_l}} \left(\mu\right)\!\right)^L\nonumber\\
\!&=\!1 \!-\! \left(\!1 \!-\! \frac{1}{e^{\lambda \mu}}\left(\!\frac{\lambda_a}{\lambda_a\!+\!\lambda \mu}\!\right)^{\alpha_a}
        \prod\nolimits_{k=1}^{K}\!\left(\frac{\lambda_k}{\lambda_k\!+\!\lambda \mu}\right)^{\alpha_k}\!\right)^L,\!\! \tag{A6}
\end{align}
establishing the desired result.
\end{IEEEproof}

\section{Proof of Theorem 4.2}
\begin{IEEEproof}
According to (60), $R_s'(\theta)$ can be calculated as
\begin{align}           
\!\!\!\!\!\!R_s'\left(\theta\right) &\!=\!  \frac{1}{\ln(2)}\!\left(  \frac{\gamma_B}{1\!+\!\theta\gamma_B} \!-\! \frac{w\left(\theta\right)\!+\!\theta w'\left(\theta\right)}{1\!+\!\theta w\left(\theta\right)}  \right) \nonumber\\
&\!=\! \frac{1}{\ln(2)}\!
\left(\!
\frac{\gamma_B\!-\!w(\theta)}{\left(1\!+\!\theta\gamma_B\right)\left[1\!+\!\theta w(\theta)\right]}
\!-\! \frac{ {w'(\theta)}/{w(\theta)}} {1\!+\! 1/\left[\theta w\left(\theta\right)\right]}
\!\right)\!. \!\!\!\!\!\tag{B1}
\end{align}
By Lemma 4.1 and Corollary 4.2, the first and second terms on the right-hand side of (B1) are monotonically decreasing and increasing with $\theta$, respectively.
Thus $R_s'\left(\theta\right)$ is a monotonically decreasing function of $\theta$, i.e., $R_{s}''(\theta)\!<\!0$.
By the second order condition \cite{Convex}, we obtain Theorem 4.2.
\end{IEEEproof}

\section{Proof of Remark 4.2}
\begin{IEEEproof}
According to SOP constraint (59),  $w\left(0^+\right)$ can be obtained by solving the following equation:
\begin{align}           
\!\!\!\!c
\!=\! \frac{w\left(0^+\right)}{P_a}\!+\!\alpha_a\!\ln\!\left(\!1\!+\!\frac{w\left(0^+\right)}{\alpha_a}\!\right)\!+\!\sum_{k=1}^{K}\alpha_k\!\ln\!\left(\!1\!+\!\frac{w\left(0^+\right)}{g_k}\!\right)\!.\!\!\tag{C1}
\end{align}
Therefore, the term $w(0^+)$ can be regarded as a threshold associated with SOP constraint (C1).
From (B1), we can obtain the following equivalences:
\begin{align}           
R_s'(0^+) \!\le\!0  \!\Leftrightarrow\! \gamma_B\!\le\!w(0^+); R_s'(0^+) \!>\!0  \!\Leftrightarrow\! \gamma_B\!>\!w(0^+). \tag{C2}
\end{align}
When $\gamma_B\!\le\!w(0^+)$, it follows by Corollary 4.3 that $R_s\!=\!0$.
When $\gamma_B\!>\!w(0^+)$, then there must exist $\theta\!\in\left(0,1\right]$ such that
$R_s\!>\!0$. Then the desired result is established.
\end{IEEEproof}

\section{Proof of Remark 4.3}
\begin{IEEEproof}
Since $\gamma_B\!=\!P_a\sigma^2\!\rightarrow\!\infty$ as $P_a\!\rightarrow\!\infty$,
it follows that $\gamma_B\!>\!w(0^+)$ is satisfied,
and thus by Remark 4.2, a positive secrecy rate is achievable and the optimal solution satisfies that $\theta^*\!>\!0$.
In addition, since $g_k\!=\!P_a\lambda_k\!\rightarrow\!\infty$ as $P_a\!\rightarrow\!\infty$ for all $k\!\in\!\mathcal{K}$,
the SOP constraint (59) can be approximated as
\begin{align}           
\!\!\!\!\ln\left(\frac{1}{1\!-\!(1\!-\!\varepsilon_{th})^{\frac{1}{L}}}\right)
\!=\! \alpha_a\ln\left(\!1\!+\!\frac{\left(1\!-\!\theta\right)w\left(\theta\right)}{\alpha_a}\!\right).\!\! \tag{D1}
\end{align}
After some rearrangement of terms, we can obtain that
\begin{align}           
&\!\!\!\!\left(1\!-\!\theta\right)w\left(\theta\right) \!=\! \alpha_a\left[\left(1\!-\!\left(1\!-\!\varepsilon_{th}\right)^{\frac{1}{L}}\right)^{\!-\!\frac{1}{\alpha_a}}\!-\!1\right]  \nonumber\\
&\!\!\!\!\Rightarrow w'\left(\theta\right)\!=\!\frac{w\left(\theta\right)}{1\!-\!\theta}\!=\!\frac{\alpha_a}{\left(1\!-\!\theta\right)^2} \left[\left(1\!-\!\left(1\!-\!\varepsilon_{th}\right)^{\frac{1}{L}}\right)^{\!-\!\frac{1}{\alpha_a}}\!-\!1\right].\!\! \tag{D2}
\end{align}
Besides, as $P_a\!\rightarrow\!\infty$, $R_s'\left(\theta\right)$ in (B1) can be approximated as
\begin{align}           
R_s'\left(\theta\right)  
\!=\!\frac{1}{\ln\left(2\right)}\frac{1\!-\!\theta^2 w'\left(\theta\right)}{\theta\left[1\!+\!\theta w\left(\theta\right)\right]}.\tag{D3}
\end{align}
It can be shown from (D2) that $w'\left(\theta\right)\!\rightarrow\!\infty$ as $\phi\!\rightarrow\!1$, and thus by (D3), $R_s'\left(1\right)\!<\!0$ holds.
From Theorem 4.2, we conclude that there exists a unique optimal solution $\theta^*\!\in\!\left(0,1\right)$ such that $R_s'\left(\theta^*\right)\!=\!0$,
which is equivalent to $1\!-\!\left(\theta^*\right)^2 {w'\left(\theta^*\right)}\!=\!0$.
Then by (D2) it follows that
\begin{align}
\alpha_a\left(\frac{\theta^*}{1\!-\!\theta^*}\right)^2 \left[\left(1\!-\!\left(1\!-\!\varepsilon_{th}\right)^{\frac{1}{L}}\right)^{\!-\!\frac{1}{\alpha_a}}\!-\!1\right]\!=\!1,\tag{D4}
\end{align}
By solving (D4), we can obtain the expression in (64).
\end{IEEEproof}



\ifCLASSOPTIONcaptionsoff
  \newpage
\fi

\bibliographystyle{IEEEtran}
\bibliography{IEEEabrv,TSP}

\end{document}